\documentclass[preprint,12pt]{elsarticle}

\usepackage[hyphens]{url}
\usepackage[hyperindex,breaklinks, allcolors=black, colorlinks=true]{hyperref}
\usepackage{graphicx}   
\usepackage{natbib}      
\usepackage{amsfonts}
\usepackage{amsmath} 
\usepackage{amssymb}
\usepackage{enumerate}
\usepackage{mathrsfs}
\usepackage{comment}

\usepackage[page]{appendix}

\usepackage{dsfont}
\usepackage{epstopdf}
\usepackage{dashrule}
\usepackage{array}
\usepackage{subfig}
\usepackage{pgffor}

\newcommand{\pda}{\pi}
\newcommand{\quant}{Q}
\newcommand{\pshort}{\pi^-}
\newcommand{\pover}{\pi^+}

\usepackage{float}

\newtheorem{theore}{Theorem}
\newtheorem{corollar}{Corollary}
\newtheorem{lemm}{Lemma}
\newtheorem{definitio}{Definition}
\newtheorem{proble}{Problem}
\newtheorem{assumptio}{Assumption}
\newtheorem{propert}{Property}
\newtheorem{propositio}{Proposition}
\newtheorem{conjectur}{Conjecture}
\newtheorem{exampl}{Example}
\newtheorem{rul}{Rule}

\newtheorem{remar}{Remark}

\def\nn{\nonumber}
\def\beq{\begin{equation}}
\def\eeq{\end{equation}}
\def\beqs{\begin{equation*}}
\def\eeqs{\end{equation*}}
\def\balign{\begin{align}}
\def\ealign{\end{align}}
\def\bea{\begin{eqnarray}}
\def\eea{\end{eqnarray}}
\def\ba{\begin{array}}
\def\ea{\end{array}}
\def\bcenter{\begin{center}}
\def\ecenter{\end{center}}
\def\bitem{\begin{itemize}}
\def\eitem{\end{itemize}}
\def\benum{\begin{enumerate}}
\def\eenum{\end{enumerate}}
\def\bdoc{\begin{document}}
\def\edoc{\end{document}}
\def\defeq{{\stackrel{\Delta}{=}}}
\def\nn{\nonumber}

\def\eg{{e.g., \/}}
\def\etal{{\it et al. \/}}
\def\viz{{\iet viz.,\ \/}}
\def\ie{{i.e.,\ \/}}
\def\etc{{\it etc.,\ \/}}
\def\cf{{cf.,\ \/}}

\usepackage[usenames,dvipsnames,svgnames,table]{xcolor}

\def\black{\textcolor{black}}
\def\yellow{\textcolor{yellow}}
\def\green{\textcolor{green}}
\def\tcbg{\textcolor{bgrd}}
\def\magenta{\textcolor{magenta}}
\def\white{\textcolor{white}}
\def\gray{\textcolor{gray}}

\newcommand\Quote[1]{\lq\textsl{#1}\rq}
\newcommand\fr[2]{{\textstyle\frac{#1}{#2}}}

\newcommand{\Ambb}{\mathbb{A}}
\newcommand{\Bmbb}{\mathbb{B}}
\newcommand{\Cmbb}{\mathbb{C}}
\newcommand{\Dmbb}{\mathbb{D}}
\newcommand{\Embb}{\mathbb{E}}
\newcommand{\Fmbb}{\mathbb{F}}
\newcommand{\Gmbb}{\mathbb{G}}
\newcommand{\Hmbb}{\mathbb{H}}
\newcommand{\Imbb}{\mathbb{I}}
\newcommand{\Jmbb}{\mathbb{J}}
\newcommand{\Kmbb}{\mathbb{K}}
\newcommand{\Lmbb}{\mathbb{L}}
\newcommand{\Mmbb}{\mathbb{M}}
\newcommand{\Nmbb}{\mathbb{N}}
\newcommand{\Ombb}{\mathbb{O}}
\newcommand{\Pmbb}{\mathbb{P}}
\newcommand{\Qmbb}{\mathbb{Q}}
\newcommand{\Rmbb}{\mathbb{R}}
\newcommand{\Smbb}{\mathbb{S}}
\newcommand{\Tmbb}{\mathbb{T}}
\newcommand{\Umbb}{\mathbb{U}}
\newcommand{\Vmbb}{\mathbb{V}}
\newcommand{\Wmbb}{\mathbb{W}}
\newcommand{\Xmbb}{\mathbb{X}}
\newcommand{\Ymbb}{\mathbb{Y}}
\newcommand{\Zmbb}{\mathbb{Z}}

\newcommand{\Amsc}{\mathcal{A}}
\newcommand{\Bmsc}{\mathcal{B}}
\newcommand{\Cmsc}{\mathcal{C}}
\newcommand{\Dmsc}{\mathcal{D}}
\newcommand{\Emsc}{\mathcal{E}}
\newcommand{\Fmsc}{\mathcal{F}}
\newcommand{\Gmsc}{\mathcal{G}}
\newcommand{\Hmsc}{\mathcal{H}}
\newcommand{\Imsc}{\mathcal{I}}
\newcommand{\Jmsc}{\mathcal{J}}
\newcommand{\Kmsc}{\mathcal{K}}
\newcommand{\Lmsc}{\mathcal{L}}
\newcommand{\Mmsc}{\mathcal{M}}
\newcommand{\Nmsc}{\mathcal{N}}
\newcommand{\Omsc}{\mathcal{O}}
\newcommand{\Pmsc}{\mathcal{P}}
\newcommand{\Qmsc}{\mathcal{Q}}
\newcommand{\Rmsc}{\mathcal{R}}
\newcommand{\Smsc}{\mathcal{S}}
\newcommand{\Tmsc}{\mathcal{T}}
\newcommand{\Umsc}{\mathcal{U}}
\newcommand{\Vmsc}{\mathcal{V}}
\newcommand{\Wmsc}{\mathcal{W}}
\newcommand{\Xmsc}{\mathcal{X}}
\newcommand{\Ymsc}{\mathcal{Y}}
\newcommand{\Zmsc}{\mathcal{Z}}

\newcommand{\Ascr}{\mathscr{A}}
\newcommand{\Bscr}{\mathscr{B}}
\newcommand{\Cscr}{\mathscr{C}}
\newcommand{\Dscr}{\mathscr{D}}
\newcommand{\Escr}{\mathscr{E}}
\newcommand{\Fscr}{\mathscr{F}}
\newcommand{\Gscr}{\mathscr{G}}
\newcommand{\Hscr}{\mathscr{H}}
\newcommand{\Iscr}{\mathscr{I}}
\newcommand{\Jscr}{\mathscr{J}}
\newcommand{\Kscr}{\mathscr{K}}
\newcommand{\Lscr}{\mathscr{L}}
\newcommand{\Mscr}{\mathscr{M}}
\newcommand{\Nscr}{\mathscr{N}}
\newcommand{\Oscr}{\mathscr{O}}
\newcommand{\Pscr}{\mathscr{P}}
\newcommand{\Qscr}{\mathscr{Q}}
\newcommand{\Rscr}{\mathscr{R}}
\newcommand{\Sscr}{\mathscr{S}}
\newcommand{\Tscr}{\mathscr{T}}
\newcommand{\Uscr}{\mathscr{U}}
\newcommand{\Vscr}{\mathscr{V}}
\newcommand{\Wscr}{\mathscr{W}}
\newcommand{\Xscr}{\mathscr{X}}
\newcommand{\Yscr}{\mathscr{Y}}
\newcommand{\Zscr}{\mathscr{Z}}

\def\Atiny{\mbox{\tiny{A}}}
\def\Btiny{\mbox{\tiny{B}}}
\def\Ctiny{\mbox{\tiny{C}}}
\def\Dtiny{\mbox{\tiny{D}}}
\def\Etiny{\mbox{\tiny{E}}}
\def\Ctiny{\mbox{\tiny{C}}}
\def\Ftiny{\mbox{\tiny{F}}}
\def\Gtiny{\mbox{\tiny{G}}}
\def\Htiny{\mbox{\tiny{H}}}
\def\Itiny{\mbox{\tiny{I}}}
\def\Jtiny{\mbox{\tiny{J}}}
\def\Ktiny{\mbox{\tiny{K}}}
\def\Ltiny{\mbox{\tiny{L}}}
\def\Mtiny{\mbox{\tiny{M}}}
\def\Ntiny{\mbox{\tiny{N}}}
\def\Otiny{\mbox{\tiny{O}}}
\def\Ptiny{\mbox{\tiny{P}}}
\def\Qtiny{\mbox{\tiny{Q}}}
\def\Rtiny{\mbox{\tiny{R}}}
\def\Stiny{\mbox{\tiny{S}}}
\def\Ttiny{\mbox{\tiny{T}}}
\def\Utiny{\mbox{\tiny{U}}}
\def\Vtiny{\mbox{\tiny{V}}}
\def\Wtiny{\mbox{\tiny{W}}}
\def\Xtiny{\mbox{\tiny{X}}}
\def\Ytiny{\mbox{\tiny{Y}}}
\def\Ztiny{\mbox{\tiny{Z}}}

\def\alphabf{\hbox{\boldmath$\alpha$\unboldmath}}
\def\betabf{\hbox{\boldmath$\beta$\unboldmath}}
\def\gammabf{\hbox{\boldmath$\gamma$\unboldmath}}
\def\deltabf{\hbox{\boldmath$\delta$\unboldmath}}
\def\epsilonbf{\hbox{\boldmath$\epsilon$\unboldmath}}
\def\zetabf{\hbox{\boldmath$\zeta$\unboldmath}}
\def\etabf{\hbox{\boldmath$\eta$\unboldmath}}
\def\iotabf{\hbox{\boldmath$\iota$\unboldmath}}
\def\kappabf{\hbox{\boldmath$\kappa$\unboldmath}}
\def\lambdabf{\hbox{\boldmath$\lambda$\unboldmath}}
\def\mubf{\hbox{\boldmath$\mu$\unboldmath}}
\def\nubf{\hbox{\boldmath$\nu$\unboldmath}}
\def\xibf{\hbox{\boldmath$\xi$\unboldmath}}
\def\pibf{\hbox{\boldmath$\pi$\unboldmath}}
\def\rhobf{\hbox{\boldmath$\rho$\unboldmath}}
\def\sigmabf{\hbox{\boldmath$\sigma$\unboldmath}}
\def\taubf{\hbox{\boldmath$\tau$\unboldmath}}
\def\upsilonbf{\hbox{\boldmath$\upsilon$\unboldmath}}
\def\phibf{\hbox{\boldmath$\phi$\unboldmath}}
\def\chibf{\hbox{\boldmath$\chi$\unboldmath}}
\def\psibf{\hbox{\boldmath$\psi$\unboldmath}}
\def\omegabf{\hbox{\boldmath$\omega$\unboldmath}}
\def\Sigmabf{\hbox{$\bf \Sigma$}}
\def\Upsilonbf{\hbox{$\bf \Upsilon$}}
\def\Omegabf{\hbox{$\bf \Omega$}}
\def\Deltabf{\hbox{$\bf \Delta$}}
\def\Gammabf{\hbox{$\bf \Gamma$}}
\def\Thetabf{\hbox{$\bf \Theta$}}
\def\Lambdabf{\mbox{$ \bf \Lambda $}}
\def\Xibf{\hbox{\bf$\Xi$}}
\def\Pibf{{\bf \Pi}}
\def\thetabf{{\mbox{\boldmath$\theta$\unboldmath}}}
\def\Upsilonbf{\hbox{\boldmath$\Upsilon$\unboldmath}}
\newcommand{\Phibf}{\mbox{${\bf \Phi}$}}
\newcommand{\Psibf}{\mbox{${\bf \Psi}$}}

\def\abf{{\bf a}}
\def\bbf{{\bf b}}
\def\cbf{{\bf c}}
\def\dbf{{\bf d}}
\def\ebf{{\bf e}}
\def\fbf{{\bf f}}
\def\gbf{{\bf g}}
\def\hbf{{\bf h}}
\def\ibf{{\bf i}}
\def\jbf{{\bf j}}
\def\kbf{{\bf k}}
\def\lbf{{\bf l}}
\def\mbf{{\bf m}}
\def\nbf{{\bf n}}
\def\obf{{\bf o}}
\def\pbf{{\bf p}}
\def\qbf{{\bf q}}
\def\rbf{{\bf r}}
\def\sbf{{\bf s}}
\def\tbf{{\bf t}}
\def\ubf{{\bf u}}
\def\vbf{{\bf v}}
\def\wbf{{\bf w}}
\def\xbf{{\bf x}}
\def\ybf{{\bf y}}
\def\zbf{{\bf z}}
\def\rbf{{\bf r}}
\def\xbf{{\bf x}}
\def\ybf{{\bf y}}
\def\Abf{{\bf A}}
\def\Bbf{{\bf B}}
\def\Cbf{{\bf C}}
\def\Dbf{{\bf D}}
\def\Ebf{{\bf E}}
\def\Fbf{{\bf F}}
\def\Gbf{{\bf G}}
\def\Hbf{{\bf H}}
\def\Ibf{{\bf I}}
\def\Jbf{{\bf J}}
\def\Kbf{{\bf K}}
\def\Lbf{{\bf L}}
\def\Mbf{{\bf M}}
\def\Nbf{{\bf N}}
\def\Obf{{\bf O}}
\def\Pbf{{\bf P}}
\def\Qbf{{\bf Q}}
\def\Rbf{{\bf R}}
\def\Sbf{{\bf S}}
\def\Tbf{{\bf T}}
\def\Ubf{{\bf U}}
\def\Vbf{{\bf V}}
\def\Wbf{{\bf W}}
\def\Xbf{{\bf X}}
\def\Ybf{{\bf Y}}
\def\Zbf{{\bf Z}}
\def\Ac{{\cal A}}
\def\Bc{{\cal B}}
\def\Cc{{\cal C}}
\def\Dc{{\cal D}}
\def\Ec{{\cal E}}
\def\Fc{{\cal F}}
\def\Gc{{\cal G}}
\def\Hc{{\cal H}}
\def\Ic{{\cal I}}
\def\Jc{{\cal J}}
\def\Kc{{\cal K}}
\def\Lc{{\cal L}}
\def\Mc{{\cal M}}
\def\Nc{{\cal N}}
\def\Oc{{\cal O}}
\def\Pc{{\cal P}}
\def\Qc{{\cal Q}}
\def\Rc{{\cal R}}
\def\Sc{{\cal S}}
\def\Tc{{\cal T}}
\def\Uc{{\cal U}}
\def\Vc{{\cal V}}
\def\Wc{{\cal W}}
\def\Xc{{\cal X}}
\def\Yc{{\cal Y}}
\def\Zc{{\cal Z}}

\def\0bf{\mathbf{0}}
\def\1bf{\mathbf{1}}

\def\lambdabf{\boldsymbol{\lambda}}
\def\mubf{\boldsymbol{\mu}}
\def\gammabf{\boldsymbol{\gamma}}
\def\xibf{\boldsymbol{\xi}}
\def\sigmabf{\boldsymbol{\sigma}}
\def\zetabf{\boldsymbol{\zeta}}

\def\Lambdabf{\boldsymbol{\Lambda}}
\def\Mubf{\boldsymbol{\Mu}}
\def\Gammabf{\boldsymbol{\Gamma}}
\def\Xibf{\boldsymbol{\Xi}}
\def\Sigmabf{\boldsymbol{\Sigma}}
\def\Zetabf{\boldsymbol{\Zeta}}

\newcommand\independent{\protect\mathpalette{\protect\independenT}{\perp}}
\def\independenT#1#2{\mathrel{\rlap{$#1#2$}\mkern2mu{#1#2}}}

\newcommand{\bsf}[1]{\textsf{\textbf{#1}}}
\newcommand{\lbsf}[1]{\textsf{\large  \textbf{#1}}}
\newcommand{\Lbsf}[1]{\textsf{\Large  \textbf{#1}}}
\newcommand{\hbsf}[1]{\textsf{\huge  \textbf{#1}}}

\newcommand{\myminipage}[3]{\begin{minipage}[#1]{#2}{#3} \end{minipage}}
\newcommand{\sbs}[4]{\myminipage{c}{#1}{#3} \hfill
\myminipage{c}{#2}{#4}}

\newcommand{\myfig}[2]{\centerline{\psfig{figure=#1,width=#2,silent=}}}
\newcommand{\myfigh}[2]{\centerline{\psfig{figure=#1,height=#2,silent=}}}
\newcommand{\myfigwh}[3]{\centerline{\psfig{figure=#1,width=#2,height=#3,silent=}}}

\newcommand{\beqa}{\begin{eqnarray}}
\newcommand{\eeqa}{\end{eqnarray}}
\newcommand{\beqan}{\begin{eqnarray*}}
\newcommand{\eeqan}{\end{eqnarray*}}
\newcommand{\dst}[1]{\displaystyle{ #1 }}
\newcounter{pf}

\newcommand{\smax}[1] { \bar \sigma \left( #1 \right) }
\newcommand{\Rn}{{\mathbb R}^n}
\newcommand{\R}{{\mathbb R}}
\newcommand{\C}{{\mathbb C}}
\newcommand{\Rm}{\mathbb{R}^m}
\newcommand{\Rmn}{\mathbb{R}^{m \times n}}
\newcommand{\Rpq}{\mathbb{R}^{p \times q}}
\newcommand{\Cn}{\mathbb{C}^n}
\newcommand{\Cm}{\mathbb{C}^m}
\newcommand{\Cnn}{\mathbb{C}^{n \times n}}
\newcommand{\Cmn}{\mathbb{C}^{m \times n}}
\newcommand{\ip}[1]{\left\langle #1 \right\rangle}
\newcommand{\rank}{\mbox{rank}}
\newcommand{\Span}{\mbox{\rm Span }}
\newcommand{\Trace}{\mbox{\rm Tr }}
\newcommand{\Spec}{\mbox{\rm Spec }}
\newcommand{\vectornorm}[1]{\left\|#1\right\|}

\newcommand{\pd}[2]{\frac{\partial #1}{\partial #2}}
\newcommand{\ppd}[3]{\frac{\partial^2 #1}{\partial #2 \partial #3}}

\newcommand{\thtilde}{\tilde{\theta}}
\newcommand{\thnom}{\theta^\circ}
\newcommand{\thopt}{\theta^{\mbox{\small opt}}}
\newcommand{\thhat}{{\hat{\theta}}}
\newcommand{\Tho}{\Theta^\circ}
\newcommand{\tho}{\theta^\circ}
\newcommand{\np}{{n_p}}

\newcommand{\ii}{{[i]}}
\newcommand{\II}{{[i+1]}}
\newcommand{\iii}{{[ii]}}
\newcommand{\jj}{{[j]}}
\newcommand{\kk}{{[k]}}
\newcommand{\thi}{{\theta^\ii}}
\newcommand{\thI}{{\theta^\II}}
\newcommand{\di}{{d^\ii}}
\newcommand{\gi}{{g^\ii}}
\newcommand{\Hi}{{\HH^\ii}}
\newcommand{\thK}{\theta^{(k+1)}}
\newcommand{\gk}{{g^{(k)}}}
\newcommand{\Hk}{{{\cal H}^{(k)}}}

\newcommand{\bfdelta}{{\bf \Delta}}

\newcommand{\pr}[1]{\mathbb{P} \left\{ #1 \right\}}
\newcommand{\prC}[1]{\mathbb{P}_0 \left\{ #1 \right\}}
\newcommand{\prN}[1]{\mathbb{P}_0^N \left\{ #1 \right\}}
\newcommand{\Exp}[1]{\exp \left\{ #1 \right\}} 
\newcommand{\gaussian}[1]{\mathbb{N} \left( #1 \right)}
\newcommand{\uniform}[1]{\mathbb{U} \left[ #1 \right]}
\newcommand{\exponential}[1]{\mathbb{E} \left[ #1 \right]}
\newcommand{\EXP}[1]{\EEXP \left[ #1 \right]} 
\newcommand{\EEXP}{\mbox{\bsf{E}}} 
\newcommand{\Prob}[1]{\mbox{{\sf Pr}} \left(#1 \right)}
\newcommand{\convas}{\stackrel{as}{\longrightarrow}}
\newcommand{\convinp}{\stackrel{p}{\longrightarrow}}
\newcommand{\convind}{\stackrel{d}{\longrightarrow}}
\newcommand{\convqm}{\stackrel{qm}{\longrightarrow}}
\newcommand{\sss}[1]{{_{#1}}}
\newcommand{\density}[2]{p_{_{_{#1}}}\!\!\left(#2 \right)} 
\newcommand{\distro}[2]{P_{_{_{#1}}}\!\!\left(#2 \right)} 
\newcommand{\rxx}[1]{R_{_{#1}}\!} 
\newcommand{\sxx}[1]{S_{_{#1}}} 
\newcommand{\cov}[1]{\Lambda_{_{#1}}} 
\newcommand{\mean}[1]{m_{_{#1}}} 
\newcommand{\LS}[1]{\hat{#1}_{_{LS}}} 
\newcommand{\MV}[1]{\hat{#1}_{_{MV}}} 
\newcommand{\LMV}[1]{\hat{#1}_{_{LMV}}} 
\newcommand{\ML}[1]{\hat{#1}_{_{ML}}} 

\renewcommand{\arraystretch}{0.9}
\newcommand{\bmat}[1]{ \begin{bmatrix} #1 \end{bmatrix}}
\newcommand{\mat}[1]{ \left[ \begin{array}{cccccccc} #1 \end{array}
\right] }
\newcommand{\smallmat}[1]{\small{\mat{#1}}}
\newcommand{\sysblk}[4]{\begin{array}{c|cccc}#1&#2\\ \hline#3&#4
\end{array}}
\newcommand{\sysmat}[4]{\left[\sysblk{#1}{#2}{#3}{#4}\right]}
\newcommand{\SGeq}{\succ}
\newcommand{\SLeq}{\prec}
\newcommand{\Geq}{\succeq}
\newcommand{\Leq}{\preceq}

\newcommand{\Bset}{\mathbb{B}}
\newcommand{\Fset}{\mathbb{F}}
\newcommand{\Mset}{\mathbb{M}}
\newcommand{\Pset}{\mathbb{P}}
\newcommand{\Rset}{\mathbb{R}}
\newcommand{\Sset}{\mathbb{S}}
\newcommand{\Tset}{\mathbb{T}}
\newcommand{\Uset}{\mathbb{U}}
\newcommand{\Vset}{\mathbb{V}}
\newcommand{\Wset}{\mathbb{W}}

\newcommand{\Ical}{{\cal I}}
\newcommand{\Acal}{{\cal A}}
\newcommand{\Bcal}{{\cal B}}
\newcommand{\Ccal}{{\cal C}}
\newcommand{\Dcal}{{\cal D}}
\newcommand{\Ecal}{{\cal E}}
\newcommand{\Fcal}{{\cal F}}
\newcommand{\Gcal}{{\cal G}}
\newcommand{\Hcal}{{\cal H}}
\newcommand{\Kcal}{{\cal K}}
\newcommand{\Lcal}{{\cal L}}
\newcommand{\Mcal}{{\cal M}}
\newcommand{\Ncal}{{\cal N}}
\newcommand{\Pcal}{{\cal P}}
\newcommand{\Qcal}{{\cal Q}}
\newcommand{\Rcal}{{\cal R}}
\newcommand{\Scal}{{\cal S}}
\newcommand{\Tcal}{{\cal T}}
\newcommand{\Wcal}{{\cal W}}
\newcommand{\Ucal}{{\cal U}}
\newcommand{\Vcal}{{\cal V}}
\newcommand{\Xcal}{{\cal X}}
\newcommand{\Zcal}{{\cal Z}}

\newcommand{\FF}{{\bf F}}
\newcommand{\GG}{{\bf G}}
\newcommand{\HH}{{\bf H}}
\newcommand{\LL}{{\bf L}}
\newcommand{\NN}{{\bf N}}
\newcommand{\MM}{{\bf M}}
\newcommand{\PP}{{\bf P}}
\newcommand{\QQ}{{\bf Q}}
\newcommand{\RR}{{\bf R}}
\renewcommand{\SS}{{\bf S}}
\newcommand{\TT}{{\bf T}}
\newcommand{\VV}{{\bf V}}
\newcommand{\WW}{{\bf W}}

\newcommand{\thk}{\theta^{(k)}}
\newcommand{\thb}{\theta^{\rm opt}}
\newcommand{\alb}{\alpha^{\rm opt}}
\newcommand{\dk}{d^{(k)}}
\newcommand{\Hinf}{{\cal H}_\infty}
\newcommand{\Htwo}{{\cal H}_2}

\renewcommand{\arraystretch}{1.1}

\newcommand{\red}[1]{{\color{red} #1}}
\newcommand{\blue}[1]{{\color{Blue} #1}}

\newcounter{l1}
\newcounter{l2}
\newcounter{l3}
\newcommand{\bdotlist}{\begin{list}{$\bullet$}{}}
\newcommand{\bboxlist}{\begin{list}{$\Box$}{}}
\newcommand{\bbboxlist}{\begin{list}{\raisebox{.005in}{{\tiny
$\blacksquare$ \ \ }}}{}}
\newcommand{\bdashlist}{\begin{list}{$-$}{} }
\newcommand{\blist}{\begin{list}{}{} }
\newcommand{\barablist}{\begin{list}{\arabic{l1}}{\usecounter{l1}}}
\newcommand{\balphlist}{\begin{list}{(\alph{l2})}{\usecounter{l2}}}
\newcommand{\bAlphlist}{\begin{list}{\Alph{l2}.}{\usecounter{l2}}}
\newcommand{\bdiamlist}{\begin{list}{$\diamond$}{}}
\newcommand{\bromalist}{\begin{list}{(\roman{l3})}{\usecounter{l3}}}

\newcommand{\ex}[1]{\begin{example} {\rm #1} \end{example}}
\newcommand{\prf}[1]{ \noindent {\em Proof:} \, #1 \hfill $\blacksquare$}

\usepackage{tikz}
\usetikzlibrary{decorations.pathreplacing}
\newcommand{\expe}[1]{\mathbb{E}^\gamma\left[#1 \right]}

\newcommand{\argmin}{\mathop{\rm argmin}}
\newcommand{\argmax}{\mathop{\rm argmax}}
\newcommand{\diag}{\mathop{\mathrm{diag}}}
\newcommand{\tr}{\mathop{\rm Tr}}
\newcommand{\conv}{\mathop{\rm conv}}
\newcommand{\var}{\mathop{\rm Var}}
\renewcommand{\b}[1]{\ensuremath{\boldsymbol{\mathrm{#1}}}}
\newcommand{\ms}{{\rm MS}}
\newcommand{\tcs}{{\rm TCS}}
\newcommand{\scs}{{\rm SCS}}

\newcommand{\E}[1]{\b{\mu}_{{#1}}}
\newcommand{\Var}[1]{{\Sigma_{#1}}}
\newcommand{\sol}[1]{\hat{x}^{#1}_N}

\def\xcap{b}
\def\ccap{\mathbf{\c}}
\def\rentftr{\Phi}
\def\rentfsr{\Sigma}
\def\rentDer{\Delta}
\def\lambdao{\lambda^{o}}

\usepackage{lipsum}
\makeatletter
\def\ps@pprintTitle{%
 \let\@oddhead\@empty
 \let\@evenhead\@empty
 \def\@oddfoot{}%
 \let\@evenfoot\@oddfoot}
\makeatother

\begin{document}

\begin{frontmatter}

\title{An Online Learning Approach to  Buying and Selling  Demand Response  \tnoteref{label1}}
 
\tnotetext[label1]{This work was supported in part by NSF grant ECCS-1351621, NSF grant IIP- 1632124, US DoE under the CERTS initiative, and the Simons Institute for the Theory of Computing. This work builds on our preliminary results,
presented at the  IFAC 2017 World Congress \citep{khezeli2017learning}. The current
manuscript differs significantly from the conference version in terms of new
results, formal proofs, and more detailed technical discussions. We thank Weixuan Lin for many helpful discussions, and his assistance in the proof of Lemma \ref{lem:delta}.}
\author[cu]{Kia Khezeli}
\ead{kk839@cornell.edu}

\author[cu]{Eilyan Bitar} 
\ead{eyb5@cornell.edu}

\address[cu]{School of Electrical and Computer Engineering, Cornell University, Ithaca, NY 14853, USA.}

\begin{abstract}
We adopt the perspective of an aggregator, which seeks to coordinate its \emph{purchase} of demand reductions from a fixed group of residential electricity customers, with its \emph{sale}  of the aggregate  demand reduction in a two-settlement wholesale energy market. The aggregator procures reductions in demand  by offering its customers a uniform price for reductions in consumption relative to their predetermined baselines.  Prior to its realization of the aggregate demand reduction, the aggregator must also determine how much energy to sell into the two-settlement energy market. In the day-ahead  market, the aggregator commits to a forward contract, which calls for the delivery of energy in the real-time market. The underlying aggregate demand curve, which relates the aggregate demand reduction to the aggregator's offered price, is assumed to be affine and subject to unobservable, random shocks. Assuming that both the parameters of the demand curve and the distribution
of the random shocks are initially unknown to the aggregator, we investigate the extent to which the aggregator might dynamically adapt its  offered prices and forward contracts to  maximize its expected profit over a  time window  of $T$ days. Specifically,  we design a dynamic pricing  and contract offering policy that resolves the aggregator's need to learn the unknown demand model with  its desire to maximize its cumulative expected profit  over time. 
In particular, the proposed pricing policy  is proven to incur a \emph{regret} over $T$ days that is no greater  than $O(\log(T)\sqrt{T})$.
\end{abstract}


\end{frontmatter}

\section{Introduction}

The large scale utilization of demand response (DR) resources has the potential to substantially improve the reliability and efficiency of electric power systems. Accordingly, several state and federal mandates have been established to facilitate the integration of  demand response resources into wholesale electricity markets. For example, FERC Order 719 mandates that Independent System  Operators (ISOs) permit  the direct sale of energy produced by DR resources into wholesale electricity markets \citep{no2008719}. 
However, as individual residential customers often posses insufficient capacity to participate in such markets directly, there  emerges the need for an intermediary, or \emph{aggregator},  with the ability to coordinate the demand response of  large numbers of residential customers for direct sale into the wholesale electricity market.  Such is consistent with the growing multitude of  ISO and utility-run DR programs, which require that aggregated DR resources have a minimum load curtailment capability. For example, the  Proxy Demand Resource (PDR) program operated by the California ISO has minimum capacity requirement of 100 kW, while the Day-Ahead Demand Response Program (DADRP) operated by the New York ISO has a more stringent capacity requirement of one MW.

In this paper, we adopt the perspective of an aggregator, which seeks to coordinate its \emph{purchase} of an aggregate demand reduction from a fixed group of residential electricity customers, with its \emph{sale}  of the aggregate  demand reduction into a two-settlement wholesale energy market.\footnote{From the perspective of the wholesale electricity market,  the provisioning of a measurable  reduction in demand from an aggregator is equivalent to an increase in supply.} Formally, this amounts to a two-sided optimization problem, which requires the aggregator to balance the cost it incurs in procuring a reduction in  demand  from participating customers against the revenue it derives from its sale of the (a priori uncertain) demand reduction into the wholesale energy  market. 

More specifically, we consider the setting in which the aggregator purchases demand reductions from  its customers using a non-discriminatory, posted price mechanism. That is to say, each participating customer
is payed for her reduction in electricity demand according to a uniform per-unit energy price determined by the aggregator. Pricing mechanisms of this form fall within the more general category of
DR programs that rely on peak time rebates (PTR) as incentives for demand
reduction. Prior to its realization of the aggregate demand reduction, the aggregator must also determine how much energy to sell into the two-settlement energy market. In the day-ahead (DA) market, the aggregator commits to a forward energy contract, which calls for delivery of the contracted energy in the real-time (RT) market. If the realized reduction in demand exceeds (falls short of) the forward contract, then the difference is sold (bought) in the RT market.  Therefore, in order to maximize its profit, the aggregator must co-optimize the DR price it offers its customers with the forward contract that it commits to in the wholesale energy market, as the former determines its ability to deliver the latter.

There are a variety of challenges that the aggregator faces in operating such DR programs. \emph{The most basic challenge is the prediction of how customers will adjust their aggregate demand in response to different DR prices}, i.e., the aggregate demand curve.  If the offered price is too low, consumers may be unwilling to curtail their demand; if the offered price is too high, the aggregator pays too much and gets more reduction than is needed. As the aggregator is initially ignorant to the customers' aggregate demand curve, the aggregator must attempt to learn a model of customer behavior over time through repeated observations of demand reductions in response to the DR prices that it offers. Simultaneously, the aggregator must jointly adjust its DR prices and forward contract offerings in such a manner as to facilitate profit maximization over time.  As we will later show, such tasks are intimately related, and give rise to a fundamental trade-off between the need to \emph{learn} (explore) and \emph{earn} (exploit).

\emph{Contribution:} \ 
In this paper, 
we study the setting in which the aggregator is faced with an aggregate demand curve that is affine in price, and subject to unobservable, additive random shocks. We assume that both the parameters of the demand curve and the probability distribution of the random shocks are fixed, but   \emph{initially unknown} to the aggregator. Faced with such ignorance, we  explore the extent to which the  aggregator might dynamically adapt its posted DR prices and offered contracts to  maximize its expected profit over a time frame of $T$ days. Specifically,  we design a causal pricing  and contract offering policy that resolves the aggregator's need to learn the unknown demand model with  its desire to maximize its cumulative expected profit  over time. 
The proposed pricing policy  is proven to exhibit \emph{regret} (relative to an oracle)  over $T$ days that  is at most $O(\log(T)\sqrt{T})$. In addition,  the proposed policy is proven to generate a sequence of posted DR prices and forward contracts that converge to the oracle optimal DR price and forward contract in the mean square sense.

\emph{Related Work:} \  There is a large body of literature in power systems concerned with the aggregation and coordination of  flexible demand-side resources to optimize certain economic objectives that an aggregator might encounter in wholesale energy or ancillary service markets. In such settings, the aggregator will typically exercise control over  the consumption of participating demand-side resources using either (1) a \emph{direct load control}  mechanism whereby the aggregator can directly regulate the consumption of  participating load resources according to a pre-specified contract  \citep{bitar2017deadline, chao1987priority, chen2014distributed, ericson2009direct,  iria2017optimal, kundu2011modeling, mathieu2015arbitraging, nayyar2016duration, sharma2013large, tan1993interruptible, xu2016demand}; or (2) an \emph{indirect load control} mechanism whereby customers  adjust their load in response to  price signals or incentives offered by the aggregator (e.g., time-of-use pricing, peak time rebates, etc.) \citep{borenstein2002dynamic, gan2013optimal, jia2014online, li2016market,  li2011optimal, ma2013decentralized, samadi2010optimal, yang2013game}. 

The literature---as it relates to the problem of co-optimizing an aggregator's (two-sided) transactions between end-use customers and the wholesale market---is much less developed.  Campaigne et al. \cite{campaigne2015firming} consider a two-sided market model  that is perhaps closest in nature to the one considered in this paper. Specifically, the authors adopt a mechanism design approach to the procurement of load reductions from customers,  where customers are rationed and remunerated according to their self-reported types.\footnote{We refer the reader to \citep{chao2012competitive, crampes2015demand} for a related line of literature, which also employs a mechanism design approach to the procurement of demand reductions in such two-sided markets.}  In this paper, we adopt a \emph{posted price} approach to the procurement of demand reductions from  customers. This is in sharp contrast to the mechanism design approach of \cite{campaigne2015firming}, as it gives rise to the need to learn customers' types (i.e., demand functions) over time from measured data. From a practical standpoint, there are a variety of reasons as to why a posted price approach might be preferable to the mechanism design approach  advocated by Campaigne et al. \citep{campaigne2015firming}, not the least of which pertains to the simplicity and ease of implementation of posted pricing schemes.  We refer the reader to \citep{einav2016auctions} for a detailed discussion surrounding the advantages and disadvantages of such an approach in the context of online marketplaces. To the best of our knowledge,  this paper is the first to analyze the use  of a posted pricing scheme by an aggregator participating in such two-sided markets.

\emph{Organization:} \ The remainder of the paper is organized as follows. In Section \ref{sec:model}, we formulate the aggregator's profit maximization problem. In Section \ref{sec:learning}, we propose a recursive estimation scheme to facilitate the online learning of the unknown demand model. In Section \ref{sec:results}, we propose an adaptive pricing and contract offering policy for the aggregator, and provide a theoretical analysis that establishes a sublinear growth rate of the regret incurred by the policy. In Section \ref{sec:numerical}, we illustrate the performance of our proposed policy with a numerical case study. A table listing the pertinent notation used in this paper can be found in the Appendix to the paper.  Detailed proofs of all formal results can be found  in the Appendix to the paper.

\section{Model} \label{sec:model}
We adopt the perspective of an aggregator who seeks  to purchase demand reductions from a fixed group of $N$ customers for sale into a two-settlement wholesale energy market. The market is assumed to repeat over multiple time periods (e.g., days) indexed by $t=1,2,\ldots$. The actions taken by the both aggregator and customers  are described in detail  in the following subsections, and concisely summarized  in Table \ref{table:timing}.

\subsection{Two-Settlement Market Model}
At the beginning of each day $t$, the aggregator commits to a forward contract for energy in the day-ahead (DA) market in the amount of $\quant_t$ (kWh). The forward contract is remunerated at the \emph{DA energy price}. The forward contract calls for delivery in the real-time (RT) market.  If the energy delivered by the aggregator (i.e., the aggregate demand reduction) falls short of the forward contract, the aggregator must purchase the shortfall in the RT market at the \emph{shortage  price}.  If the energy delivered exceeds the forward contract, the aggregator must sell the excess supply in the RT market at the \emph{overage price}.\footnote{We note that this two-settlement market structure reflects existing market rules, which  govern the behavior of aggregators in a variety of DR programs in operation today---including  the day-ahead demand response program (DADRP) and the proxy demand resource (PDR) program administered by the New York ISO and the California ISO, respectively.} Naturally, the wholesale energy prices will vary from day to day. We denote the wholesale energy prices (measured in  \$/kWh) on day $t$   by:
\begin{itemize}
\item $\pda_t$,  DA energy price,
\item $\pshort_t$, RT shortage price,
\item $\pover_t$, RT overage price.
\end{itemize}

We make several standard assumptions regarding the aggregator's actions and the determination of energy prices in the wholesale market.  First, we assume  that the aggregator's maximum demand  curtailment capacity is  small relative to the total volume of the DA energy market. Under this assumption, it is reasonable to assume that the aggregator cannot appreciably affect price. Accordingly, we assume that the aggregator behaves as a \emph{price taker} in the DA energy market,  and model the DA energy price $\pda_t$  as \emph{fixed} and \emph{known} at the outset of each period $t$. Second, as the  RT imbalance prices $(\pshort_t, \pover_t)$ are \emph{not known} to the aggregator at the time of committing to a forward contract in the DA market,  we model them as  random variables 
whose expected values are  denoted by 
$$\mu^- := \mathbb{E}[ \pshort_t] \quad  \text{and}    \quad \mu^+ := \mathbb{E}[ \pover_t]$$ for each period $t$. Note that while we allow the RT imbalance price \emph{realizations} to vary across time, we require that their \emph{expected values} be time invariant. We make the following technical assumption in a similar manner to \cite{campaigne2015firming}.
\begin{assumptio} \label{ass:price}   The DA energy price satisfies $\pda_t >0  $ and   $\mu^+<\pda_t<\mu^-$ on each day $t$.  
\end{assumptio}
 Assumption \ref{ass:price}  serves to facilitate clarity of exposition and analysis in the sequel, as it will preserve the  concavity of the aggregator's expected profit function \eqref{eq:r_t}. Moreover, this assumption eliminates the possibility of perverse market outcomes in which the aggregator offers forward energy contracts with the explicit intention of deviating from the contract in the RT market.

\subsection{Demand Response Model}
In order to fulfill its forward contract commitment $\quant_t$ on day $t$, the aggregator must elicit an aggregate reduction in demand from its customers. It does so by broadcasting a uniform  DR price $p_t \geq 0$, to which each customer $i$ responds with a reduction in demand in the amount of $D_{it}$ (kWh), thereby entitling each customer $i$ to receive a payment of $p_t D_{it}$. We note that implicit in this model is the assumption that each customer's reduction in demand is measured against a \emph{predetermined baseline.} The question as to how to accurately estimate  baseline demand 
 is a challenging and active area of research \citep{chao2011demand, chelmis2015curtailment, coughlin2009statistical, ma2017modeling}. The generalization of our model to accommodate the endogenous estimation of a priori uncertain customer baselines is left as a direction for future research.

\begin{table}[t] 
\centering 
\begin{tabular}{l c l}
\hline
\hline      \\ [-1.5ex]                 
Actor(s) & Decision & Description  \vspace{1mm}\\  
\hline    \\ [-1.5ex]   
Aggregator & $Q_t$ & In the day-ahead (DA) market on day $t$, the  \\ & & aggergrator commits to a forward energy  contract \\ & &  $Q_t$, which calls for delivery over pre-specified \\ & &  interval of time in the real-time (RT) market. \vspace{4mm}\\

Aggregator  &  $p_t$  & Prior to delivery in  the RT market, the aggregator \\ & &  broadcasts a uniform price $p_t$ for demand  reduction \\ & &  to all  customers participating in the DR program.   \vspace{4mm}\\ 

Customers & $D_t$ & In the RT market, participating customers respond \\ & & to the aggregator's offered price $p_t$  by reducing  \\ & & their aggregate  demand  by an amount $D_t$.   \vspace{1mm}\\ 
\hline   
\end{tabular}
\caption{Description and timing of actions taken by the aggregator and customers. }
\label{table:timing}

\end{table}

We model the response of  each customer $i$ to the posted price $p_t$ at time $t$ according to the \emph{affine} function
\begin{align*}
D_{it} = a_ip_t+b_i+\varepsilon_{it}, \quad \text{for} \  \  i = 1, \dots, N,
\end{align*}
where $a_i \in \Rset$ and $b_i \in \Rset$ are customer $i$'s idiosyncratic demand model parameters, and $\varepsilon_{it}$ is an unobservable demand shock, which we model as a zero-mean random variable. We assume that both the model parameters $a_i$ and $b_i$, and the probability distribution function of the demand shock are initially \emph{unknown to the aggregator}. Clearly, the aggregate demand reduction  $D_t:=\sum_{i=1}^N D_{it}$ satisfies the affine relationship
\begin{align}
D_{t} = ap_t+b+\varepsilon_{t}, \label{eq:model}
\end{align} 
where the aggregate demand model parameters and shock are defined as $a:=\sum_{i=1}^N a_i$, $b:=\sum_{i=1}^N b_i$, and $\varepsilon_{t}:=\sum_{i=1}^N \varepsilon_{it}$, respectively.  In the sequel, we will occasionally denote the tuple of aggregate  demand  parameters according to $\theta := (a,b)$.

We assume throughout the paper that $a \in\left[\underline{a},\overline{a}\right]$ and $b \in\left[0,\overline{b}\right]$, where the parameter bounds $\underline{a}$, $\overline{a}$, and $\overline{b}$ are assumed to be known and satisfy $0<\underline{a}\leq \overline{a}<\infty$ and $0\leq \overline{b}<\infty$. Such assumptions are natural, as they ensure a bounded and positive price elasticity of aggregate demand, and that reductions in aggregate demand are guaranteed to be nonnegative in the absence of demand shocks.  In addition to the following technical assumption, we also assume that the sequence of aggregate demand shocks $\{\varepsilon_t\}$ are independent and identically distributed (IID) random variables, which are mutually independent from the RT imbalance prices $\{\pover_t\} $ and $\{\pshort_t\}$.

\begin{assumptio} \label{ass:bilip} The aggregate demand shock $\varepsilon_t$ takes values in the interval $[\underline{\varepsilon}, \overline{\varepsilon}]$. Moreover, its cumulative distribution function $F$ is bi-Lipschitz over this range. Namely, there exists a real constant $L \geq 1$, such that for all $x,y\in [\underline{\varepsilon}, \overline{\varepsilon}]$, it holds that
 \begin{align*}
\frac{1}{L}\left| x-y \right| \leq \left| F(x)-F(y) \right| \leq L\left| x-y \right|.
\end{align*}
\end{assumptio}
The assumption that the aggregate demand shock takes bounded values is natural, given the  physical limitation on the range of values that demand can take. We also note that we do not require the aggregator to have explicit knowledge of the parameters specified in Assumption \ref{ass:bilip} beyond the assumption of their boundedness.

\subsection{Aggregator Profit}
The expected profit derived by the aggregator during period $t$ given a fixed forward contract $Q_t$ and price $p_t$  is determined by
\begin{align*}
r_t(\quant_t,p_t) := \pda_t\quant_t+\mathbb{E}\left[\pover_t[D_t-\quant_t]^+-\pshort_t[\quant_t-D_t]^+-p_tD_t\right],
\end{align*} 
where $[x]^+:=\max\{0,x\}$  for all $x \in \Rset$. 
Given our previous assumption that the demand shocks $\{\varepsilon_t\}$  are mutually independent from the RT imbalance prices $\{\pover_t\}$ and $\{\pshort_t\}$,  the expected profit function simplifies to
\begin{align}
r_t(\quant_t,p_t) = \pda_t\quant_t+\mu^+\mathbb{E}\left[[D_t-\quant_t]^+\right]-\mu^-\mathbb{E}\left[[\quant_t-D_t]^+\right]-\mathbb{E}\left[p_tD_t\right]. \label{eq:r_t}
\end{align}
Here, expectation is taken with respect to the random  demand shock $\varepsilon_t$.

We define the \emph{oracle optimal contract and price} as 
\begin{align} \label{eq:opt_OR}
(\quant^*_t,p^*_t):=\argmax\{ r_t(\quant,p):(Q,p)\in \Rset^2\}.
\end{align}
That is to say, $(\quant^*_t,p^*_t)$ denote the forward contract and DR price, which jointly maximize the aggregator's expected profit on day $t$ given perfect knowledge of the demand model. It is straightforward to calculate the oracle optimal contract and price from the first-order optimality condition associated with problem \eqref{eq:opt_OR}, as
the expected profit criterion \eqref{eq:r_t} is guaranteed to be jointly \emph{concave} in its arguments $(Q_t,p_t)$ given that satisfaction of Assumption \ref{ass:price}. The closed-form expressions for the oracle optimal contract and price are given in the following lemma. 
\begin{lemm}[Oracle Optimal Policy] \label{lem:oracle} For each period $t \geq 1$, 
the oracle optimal contract $Q^*_t$ and price $p^*_t$ are given by 
\begin{align}
\quant^*_t&=\frac{1}{2}\left(a\pda_t+b\right)+F^{-1}(\alpha_t)\label{pol:oracle q},\\
p^*_t&=\frac{1}{2}\left(\pda_t-\frac{b}{a}\right)\label{pol:oracle p},
\end{align}
where $$\alpha_t:= \frac{\pda_t-\mu^+}{\mu^- - \mu^+}.$$
\end{lemm}
Here, $F^{-1}(\alpha_t) := \inf\{x\in\Rset : F(x)\geq \alpha_t\} $  denotes the  $\alpha_t$-quantile of the random demand shock $\varepsilon_t$.  Assumption \ref{ass:price} ensures that the price ratio $\alpha_t$ is a valid probability, i.e., $\alpha_t \in (0,1)$. It is also worth noting that the oracle optimal contract can be equivalently rewritten as $\quant^*_t=ap^*_t+b+F^{-1}(\alpha_t)$. It follows that $\quant^*_t$ can be interpreted as the maximum  demand reduction that the aggregator is guaranteed to receive with probability at least $1-\alpha_t$ under the oracle optimal price $p_t^*$.

\begin{remar}[Supply Function Offer] \label{rem:supply} The oracle optimal contract $\quant^*_t$ can be equivalently  interpreted as a \emph{supply function offer} in the DA market, indicating  the maximum amount of energy that the aggregator is willing to supply at  a given price $\pi_t$. In particular, it is not difficult to show that the oracle optimal contract $\quant^*_t$ is a monotone, non-decreasing function in the DA price $\pi_t$.   Of primary importance to this interpretation is the assumption that the aggregator behaves as a price taker in the DA market---ensuring that it wields no influence over the DA market price.
\end{remar}

We define the \emph{oracle optimal profit} accumulated over $T$ time periods as
\begin{align*}
R^*(T):=\sum_{t=1}^T r_t(\quant^*_t,p^*_t).
\end{align*}
We employ the term \emph{oracle}, as $R^*(T)$ equals the maximum expected profit that an aggregator might derive over $T$ times periods if it had  perfect knowledge of the demand model at the outset.

\subsection{Policy Design and Regret}
We consider the scenario in which the aggregator knows neither the demand model parameter $\theta = (a,b)$ nor the aggregate shock distribution $F$ at the outset. Accordingly, the aggregator must endeavor to learn these features from the data that it collects over time, e.g., through online assimilation of measurements of aggregate demand reductions  in response to its posted DR prices. At the same time, the aggregator must dynamically adapt its sequence of posted DR prices (and forward contract offerings) to improve its profit over time.  In what  follows, we describe the space of  feasible policies that the aggregator might use to guide its adaptation of contracts $\{Q_t\}$ and DR prices $\{p_t\}$ over time.

Prior to its determination of the contract $\quant_t$ and the price $p_t$ at time $t$, the aggregator has access to the entire history of prices, contract offerings, and aggregate demand reductions, up to and including time period $t-1$. We define a \emph{feasible policy} as an infinite sequence of functions $\gamma:=((\quant_1,p_1),(\quant_2,p_2),\ldots)$, where each function in the sequence is allowed to depend only on the past data available until that point in time. More formally, we require that the functions $(\quant_t,p_t)$  be measurable according to the $\sigma$-algebra generated by the history of offered contracts, prices, and demand observations, i.e., \[(\quant_1,\ldots,\quant_{t-1},p_1,\ldots,p_{t-1},D_1,\ldots,D_{t-1})\] 
for all time periods $ t\geq 2$.  For the initial time period $t=1$, we require that $(\quant_1,p_1)$ be a pair of deterministic constants, as the aggregator has yet to collect any information about demand.

The \emph{expected profit} generated by  a feasible policy $\gamma$  over $T$ time periods is defined as 
\begin{align}
R^\gamma(T):=\mathbb{E}^\gamma\left[\sum_{t=1}^T r_t(\quant_t,p_t)\right], \label{eq:R_T}
\end{align}
where the expectation is taken with respect to the demand model \eqref{eq:model} under the policy $\gamma$. We measure the performance of a feasible policy $\gamma$  over $T$ time periods according to the $T$-\emph{period regret}, which is defined as
\begin{align*}
\Delta^\gamma(T):=R^*(T)-R^\gamma(T).
\end{align*}
The $T$-period regret incurred by a feasible policy equals the difference between the oracle optimal profit and the expected profit incurred by that policy over $T$ time periods. Clearly, policies that produce low regret are preferred, as the oracle optimal profit is an upper bound on the maximum expected profit achievable by any feasible policy. Accordingly, we seek the design of policies whose $T$-period regret grows sublinearly with the horizon $T$. Such policies are said to have \emph{no-regret} in the long run, as their average regret $(1/T)\cdot \Delta^\gamma(T)$  is guaranteed to vanish asymptotically. More formally, we have the following definition.
\begin{definitio}[No-Regret Policy]
A feasible policy $\gamma$ is said to have \emph{no-regret} if $\lim_{T\rightarrow\infty}\Delta^\gamma(T)/T=0$.
\end{definitio}

The following result establishes an upper bound on the $T$-period regret in terms of the  pricing and contract errors relative to their oracle optimal counterparts. Lemma \ref{lem:delta} will prove useful to the derivation of our main results.
\begin{lemm}\label{lem:delta}
The $T$-period regret incurred by any feasible policy $\gamma$ is upper bounded by
\begin{align}
&\Delta^\gamma (T)\leq a\sum_{t=1}^T \Embb^\gamma \left[ (p_t - p^*_t)^2 \right] + L(\mu^- - \mu^+) \sum_{t=1}^T \Embb^\gamma \left[\left( \quant_t-\quant^*_t-a (p_t-p^*_t) \right)^2 \right], \label{eq:delta bound}
\end{align}
where $(Q^*_t,p^*_t)$ denote the oracle optimal contract and price at time $t$.
\end{lemm} 
Lemma \ref{lem:delta} reveals that convergence of the offered contracts and posted prices to their oracle optimal counterparts (in the mean square sense) will prove essential to the design of policies that exhibit no-regret. In the following section, we introduce a simple method for demand model learning based on least-squares estimation that will facilitate the design of such policies. 

\section{Demand Model Learning} \label{sec:learning}
In this section,  we propose a simple approach to enable the dynamic learning of the underlying 
demand model from data using the method of least squares estimation.

\subsection{Parameter Estimation} \label{sec:paramerror}
We define the \emph{least squares estimator} (LSE) of the parameter $\theta$, given the history of past  prices and demand observations through time period $t$ as
\begin{align*}
\theta_t := {\arg \min} \left\{ \sum_{k=1}^{t} \left(D_k -  (\vartheta_1 p_k+\vartheta_2)\right)^2:(\vartheta_1,\vartheta_2)\in \Rset^2 \right\},
\end{align*}
for time periods $t= 2,3,\ldots$. The LSE is given by
\begin{align}
\theta_t = P_{t}^{-1} \left(  \sum_{k=1}^{t} \bmat{p_k \\ 1} D_k \right), \label{eq:theta_t}
\end{align}
assuming that the indicated inverse exists. The matrix $P_{t}$ is defined as 
\begin{align*}
P_{t}:= \sum_{k=1}^{t} \bmat{p_k \\ 1} \bmat{p_k \\ 1}^\top.
\end{align*}
Its inverse is given by
\begin{align}
P_t^{-1}=   \frac{1}{t^2V_t}\left(\sum_{k=1}^{t} \bmat{-1\\p_k} \bmat{-1\\p_k}^\top\right), \label{eq:J}
\end{align}
where $V_t:=(1/t)\sum_{k=1}^{t}(p_k-\bar{p}_{t})^2$ denotes the  \emph{sample variance} associated with the sequence of posted prices through time period $t$, and  $\bar{p}_t:=(1/t)\sum_{k=1}^{t}p_k$ denotes their \emph{sample mean}. 
The parameter estimation error that  results under the LSE  \eqref{eq:theta_t} can be expressed as
\begin{align}\label{eq:LSE}
\theta_t -\theta = P_{t}^{-1} \left(  \sum_{k=1}^{t} \bmat{p_k \\ 1} \varepsilon_k \right).
\end{align}
The expression for the parameter estimation error in \eqref{eq:LSE} hints at a dependency between the rate at which the parameter estimation error converges to zero, and the rate at which the  variance in the underlying sequence of posted prices grows (or decays) over time. In Section \ref{sec:results}, we leverage on this insight to design a pricing policy that generates \emph{enough variance} in the sequence of posted prices to ensure convergence of the sequence of parameter estimates to the true parameters in the mean square sense.

We close this section by  recalling our previous assumption that the unknown parameter $\theta$ belongs to a closed and compact set given by $\Theta := [\underline{a}, \overline{a}] \times [0, \overline{b}]$. Using this assumption, one can improve upon the LSE \eqref{eq:theta_t}   by projecting $\theta_t$ onto the set $\Theta$. More precisely, define the \emph{truncated least squares estimator} (TLSE) as
\begin{align} \label{eq:TLSE}
\widehat{\theta}_t := \arg \min \left\{  \| \vartheta - \theta_t \|_2    :  \vartheta \in \Theta \right\}.
\end{align}
It clearly holds that $\|\widehat{\theta}_t-\theta\|\leq \|\theta_t-\theta\|$, i.e., the TLSE is no worse than the LSE.


\subsection{Quantile Estimation}
We propose an approach to the recursive estimation of the unknown quantile function using the estimation residuals generated by the truncated LSE \eqref{eq:TLSE}. At each time $t$, define the sequence of \emph{residuals} associated with the estimator $\widehat{\theta}_t$ as 
\begin{align}
\widehat{\varepsilon}_{k,t} :=  D_k - (\widehat{a}_tp_k+\widehat{b}_t),\quad  \text{for} \ k=1,\dots, t. \label{eq:res}
\end{align}
Define their \emph{empirical distribution function} as 
\begin{align*}
\widehat{F}_{t}(x) : = \frac{1}{t}\sum_{k=1}^{t}\mathds{1}{\left\{\widehat{\varepsilon}_{k,t}\leq x\right\}},
\end{align*}
and their corresponding \emph{empirical quantile function} as $\widehat{F}_{t}^{-1}(\alpha) : =  \inf \{x \in \Rset : \widehat{F}_{t}(x) \geq \alpha\}$ for all $\alpha \in (0,1)$. 
It will prove useful to the subsequent analyses to express the empirical quantile function in terms of the order statistics associated with the sequence of residuals. The \emph{order statistics} associated with the sequence $\widehat{\varepsilon}_{1,t},\ldots,\widehat{\varepsilon}_{t,t}$ are defined as a permutation of the sequence denoted by $\widehat{\varepsilon}_{(1),t},\ldots,\widehat{\varepsilon}_{(t),t}$, where $$\widehat{\varepsilon}_{(1),t}\leq\widehat{\varepsilon}_{(2),t}\leq  \ldots\leq \widehat{\varepsilon}_{(t),t}.$$ With the  order statistics of the residuals in hand, one can express the empirical quantile function as
\begin{align}
\widehat{F}^{-1}_{t}(\alpha)=\widehat{\varepsilon}_{(i),t},\label{eq:order stat}
\end{align}
where $i$ is the unique index  that satisfies $i-1<t\alpha\leq i$. It is not difficult to show that this index is given by $i=\lceil t\alpha \rceil$. Using Equation \eqref{eq:order stat}, the quantile estimation error can be linked to the parameter estimation error via the following inequality,
\begin{align}
&| \widehat{F}_{t}^{-1}(\alpha)  - F^{-1}(\alpha)| \leq  | F_{t}^{-1}(\alpha)  - F^{-1}(\alpha) |  + \sqrt{1 + p_{(i)}^2} \| \widehat{\theta}_t - \theta \|,  \label{eq:quant}
\end{align}
where $F_{t}^{-1}$ is defined as the  empirical quantile function associated with sequence of demand shocks $\varepsilon_1, \dots,\varepsilon_{t}$.  

It follows from the inequality in \eqref{eq:quant} that  consistency of the quantile estimator \eqref{eq:order stat} depends on consistency of both the parameter estimator $\widehat{\theta}_t$ and the empirical quantile function  $F_t^{-1}$.  We establish consistency of the parameter estimator under our proposed policy in Lemma \ref{lem:mse}.  Clearly, consistency of the empirical quantile function $F_t^{-1}$ does not depend on the particular policy being used. In Proposition \ref{prop:shock quantile}, we establish a bound on the rate at which the sequence of functions $\{F_t^{-1}\}$ converges pointwise in probability to $F^{-1}$ on the interval $(0,1)$.

\begin{propositio}
\label{prop:shock quantile} 
There exists a  finite  positive constant $\mu_1$ such that 
\begin{align}
\mathbb{P}\{| {F}_{t}^{-1}(\alpha)  - F^{-1}(\alpha) | >  \delta \}&\leq 2\exp(-\mu_1 \delta^2t)\label{eq:dvoretzky}
\end{align}
for all $\alpha \in (0,1)$, $\delta >0$,  and $t\geq 2$.
\end{propositio}
We omit a formal proof of Proposition \ref{prop:shock quantile}, as it can be obtained as a direct consequence of Lemma 2 in \cite{dvoretzky1956asymptotic}  using Assumption \ref{ass:bilip} in this paper.

\section{Learning to Buy and Sell with No-Regret}\label{sec:results}

In what follows, we build on the approach to demand model learning  outlined in 
Section \ref{sec:learning} to construct a pricing and contract offering policy, which is
guaranteed to exhibit \emph{no-regret}. In doing so, we establish in Theorem \ref{thm:regret} a $O(\log(T)\sqrt{T})$ upper bound on the  $T$-period regret incurred under the  policy that we propose.

\subsection{Myopic Policy (MP)}
We first introduce a natural approach to pricing and contract offering, which combines the model learning scheme outlined in Section \ref{sec:learning} with a natural, albeit \emph{myopic}, approach to pricing and contract offering. That is to say, at each time period $t$, the aggregator estimates the demand model parameters and quantile function according to $\widehat{\theta}_{t-1}$ and $\widehat{F}_{t-1}^{-1}(\alpha)$ defined in \eqref{eq:TLSE} and \eqref{eq:order stat},
respectively, and sets the forward contract and price according to
\begin{align}
\widehat{\quant}_{t}&=\frac{1}{2}\left(\widehat{a}_{t-1}\pi_t+\widehat{b}_{t-1}\right)+\widehat{F}_{t-1}^{-1}(\alpha_t)\label{pol:myopic q},\\
\widehat{p}_{t}&=\frac{1}{2}\left(\pda_t-\frac{\widehat{b}_{t-1}}{\widehat{a}_{t-1}}\right)\label{pol:myopic p}.
\end{align} 
Under this myopic policy,\footnote{It is worth noting that, in the adaptive control theory literature, such myopic policies are more commonly known as \emph{certainty equivalent} policies.} the aggregator  treats its demand model
estimates in each period as if they were correct, and ignores
the  impact that  its choice of price might have on its ability to
accurately estimate the demand model in future time periods. As discussed in Section \ref{sec:paramerror}, consistency of the parameter estimator is reliant upon sufficient variance in the underlying sequence of prices. However, under the myopic policy the sequence of prices may converge prematurely to a fixed price, which differs from the oracle optimal price. As a consequence, the sequence of parameter estimates may also converge to a value that is different from the true model parameter. This phenomenon---also known as \emph{incomplete learning}---is well-documented in the adaptive control literature \citep{borkar1982identification, kumar2015stochastic, lai1982iterated} and the  revenue management  literature \citep{den2013simultaneously, keskin2014dynamic}. In Section \ref{sec:numerical}, we conduct a numerical case study, which suggests the occurrence of incomplete learning under the myopic policy. We refer the reader to Figure \ref{fig:sequence}(c) for a graphical illustration of incomplete learning under the myopic policy.

\subsection{Randomly Perturbed Myopic Policy (RPMP)}
To prevent the  occurrence of incomplete learning, we propose a novel policy that is guaranteed to generate adequate price dispersion through application of random perturbations to the myopic policy. We refer to this policy as the \emph{randomly perturbed myopic policy} (RPMP). We initialize the RPMP with a deterministic choice of prices and contracts for periods one and two $p_1$, $Q_1$, $p_2$, and $Q_2$, respectively, such that $p_1\neq p_2$.\footnote{This condition is necessary to ensure invertibility of the matrix $P_2$.} For all subsequent time periods $t \geq 3$, the RPMP sets prices and contracts according to:
\begin{align}
\quant_{t} &= \widehat{\quant}_t,\label{pol:perturbed q_1}\\[0.5em]
p_{t}      &= \begin{cases}
\widehat{p}_{t},& \text{if}\ \xi_t=0,\\
\bar{p}_{t-1}+\rho,& \text{if}\ \xi_t=1,
\end{cases}\label{pol:perturbed p_1}
\end{align}
where $\bar{p}_{t-1} = \frac{1}{t-1} \sum_{k=1}^{t-1} p_k$ denotes the sample mean of the posted price history.
Here,  $$\xi_t\sim \mbox{Ber}(\eta t^{-r})$$  defines a sequence of independent Bernoulli random variables with probabilities $\mathbb{P}\{\xi_t = 1\} = \eta t^{-r}$. The parameters   $\eta  \in (0,1]$,  $\rho \in (0, \infty)$,   and  $r \in [0, \infty )$ are user specified constants. The parameter $\eta$ determines, in part, the probability that a price perturbation is applied at any given time period, while the parameter $\rho$ determines the magnitude of this perturbation. In this paper, we allow the parameters $\eta$ and $\rho$ to be arbitrary,  and investigate the role that the parameter $r$ plays in controlling the rate at which the perturbation probability  decays over time.   We refer the reader to the discussion immediately following Lemma \ref{lem:mse} for a precise explanation of the role that the parameter $r$ plays in controlling the degree to which the  randomly perturbed myopic policy (RPMP) 
balances  \emph{exploration} versus \emph{exploitation.} It also important to note that although the parameters  $\eta$ and $\rho$ play a role in determining the  performance of the randomly perturbed myopic policy (RPMP) in \emph{finite-time}, they do not affect asymptotic performance of the policy, i.e., the   asymptotic order of regret incurred under the RPMP remains unchanged  for any choice of $\eta  \in (0,1]$ and  $\rho \in (0, \infty)$.

In the following Lemma, we establish an upper bound on the mean squared error (MSE) of the TLSE under the RPMP, which we will subsequently use to derive our main result.
\begin{lemm}[Consistency of TLSE] \label{lem:mse}
\label{lem:pvariance} Let $r\in(0,1)$. There exists a finite positive constant $K$ such that mean-squared parameter estimation error incurred under  the randomly perturbed perturbed myopic policy (RPMP) \eqref{pol:perturbed q_1}-\eqref{pol:perturbed p_1} is upper bounded by
\begin{align*}
\mathbb{E}^{\gamma}\left[\|\widehat{\theta}_t-\theta\|^2\right]\leq K\frac{\log(t)}{t^{1-r}}
\end{align*}
for all $t\geq 3$.
\end{lemm}
This characterization of mean-squared parameter estimation error will play a central role in the proof of Theorem \ref{thm:regret}, which establishes an $O(\log(T)T^r \vee T^{1-r})$ upper bound on the $T$-period regret incurred by the randomly perturbed myopic policy.

Ultimately, the parameter $r$  must  be designed to balance a delicate tradeoff between exploration and exploitation.  
On the one hand, the probability that a perturbation occurs should decay at a rate that is \emph{slow enough} to generate sufficient price dispersion necessary to ensure consistent parameter estimation (cf. Lemma \ref{lem:pvariance}). On the other hand,  this perturbation probability should decay at a rate that is \emph{fast enough} to ensure that the  (deliberate) pricing errors do not accumulate too rapidly.  In Theorem \ref{thm:regret}, we establish an upper bound on the $T$-period regret that captures this tradeoff, and show that a perturbation probability $\mathbb{P}\{\xi_t = 1\} = O(t^{-1/2})$ (i.e., $r=1/2$) is `optimal' in the sense that it minimizes the asymptotic order of our upper bound on regret up to a multiplicative logarithmic factor.

\subsection{A Bound on Regret}
In what follows, we establish an upper bound on the $T$-period regret incurred by the randomly perturbed myopic policy. As part of our main result in Theorem \ref{thm:regret},  we also characterize the optimal `decay rate' for the perturbation probability.

\begin{theore}[Sub-linear Regret]\label{thm:regret} Let $ r \in (0,1)$.
There exist finite positive constants $C_0$, $C_1$, and $C_2$ such that the  $T$-period regret  incurred under the randomly perturbed myopic policy  \eqref{pol:perturbed q_1}-\eqref{pol:perturbed p_1}  is upper bounded by
\begin{align*}
\Delta^\gamma(T)\leq C_0 + C_1\log(T) + \left(\frac{C_2}{1-r}\right)T^{1-r} + \left(\frac{C_2}{r}\right)\log(T)T^r 
\end{align*}
for all $T \geq 3$.
\end{theore}

The structure of the upper bound on regret in Theorem \ref{thm:regret} reveals an explicit \emph{exploration-exploitation} trade-off in choosing the dispersion parameter $r$. Specifically, the $O(\log(T)T^r)$ term captures the component of revenue loss driven by the parameter estimation error; and  the $O(T^{1-r})$  term  captures the component  of revenue loss driven by the  deliberate pricing errors that are incurred when price perturbations are applied.  A smaller (larger)  value of  the dispersion parameter $r$ implies a greater tendency towards exploration (exploitation) in pricing under the RPMP. Clearly, this exploration-exploitation trade-off is balanced by setting the dispersion parameter to $r= 1/2$, as this value minimizes the asymptotic order of our upper bound on regret  (up to a multiplicative logarithmic factor), yielding $$\Delta^\gamma(T)\leq O(\log(T)\sqrt{T}).$$

We note that as part of the proof of Theorem \ref{thm:regret}, we also show that the posted price sequence $\{p_t\}$ and contract sequence $\{\quant_t\}$ generated by the randomly perturbed myopic policy converge in the mean square sense to the oracle optimal price sequence $\{p_t^*\}$ and contract sequence $\{\quant_t^*\}$, respectively. It is also worth noting that Chen et al.  \cite{chen2014coordinating} consider a related setting, which entails the online control of a dynamic inventory system through pricing and ordering decisions. They consider a different class of policy designs, and establish an $O(\sqrt{T})$ upper bound on the order of regret for the class of policies they consider.

\section{Numerical Case Study} \label{sec:numerical}

We compare the performance of the myopic policy (MP) against the randomly perturbed myopic policy (RPMP) over a time horizon of $T=2500$  periods. We set the tuning parameters of the RPMP  as $\eta=0.2$, $\rho=0.08$, and $r = 0.5$. This choice of $\rho$ amounts to increasing the average   DR price offered to customers by eight cents anytime a perturbation is applied.  We assume that there are $N=10^4$ customers participating in the DR program.  For each customer $i$, we select $a_i$ uniformly at random from the interval $[0.04, 0.20]$, and independently select $b_i$ according to an exponential distribution (with mean equal to $0.01$) truncated over the interval $[0,0.1]$. This range of parameter values  is consistent with the range of  demand price elasticities observed in several real-time pricing programs operated in the United States  \citep{qdr2006benefits,faruqui2010household}.  We further assume that the idiosyncratic demand parameters are drawn independently across customers. For each customer $i$, we let the demand shock have a zero-mean normal distribution with  standard deviation equal to $0.5$, truncated over the interval $[-2,2]$. We set the DA energy price, the mean RT shortage price, and the mean RT overage price to $\pda_t  = 0.5$ (for all $t$), $\mu^- =1.7$, and $\mu^+ =0.2$  (\$/kWh), respectively. We initialize both the MP and the RPMP with a choice of prices and contracts for periods one and two $p_1=0$, $Q_1=0$, $p_2=0.25$, and $Q_2=0$, respectively. Finally, we
 estimate the  empirical means and confidence intervals associated with price, contract, and parameter estimate trajectories using 100 independent realizations of the experiment.

\begin{figure}[H]
\centering
\subfloat{\raisebox{0.11\columnwidth}[0pt][0pt]{ \makebox[0.059\columnwidth]{\small $\underset{\text{(\$/kWh)}}{p_t}$}}} 
\hspace{0.02\columnwidth}
\subfloat{\label{fig:p_t}\fbox{\includegraphics[width=0.29\columnwidth]{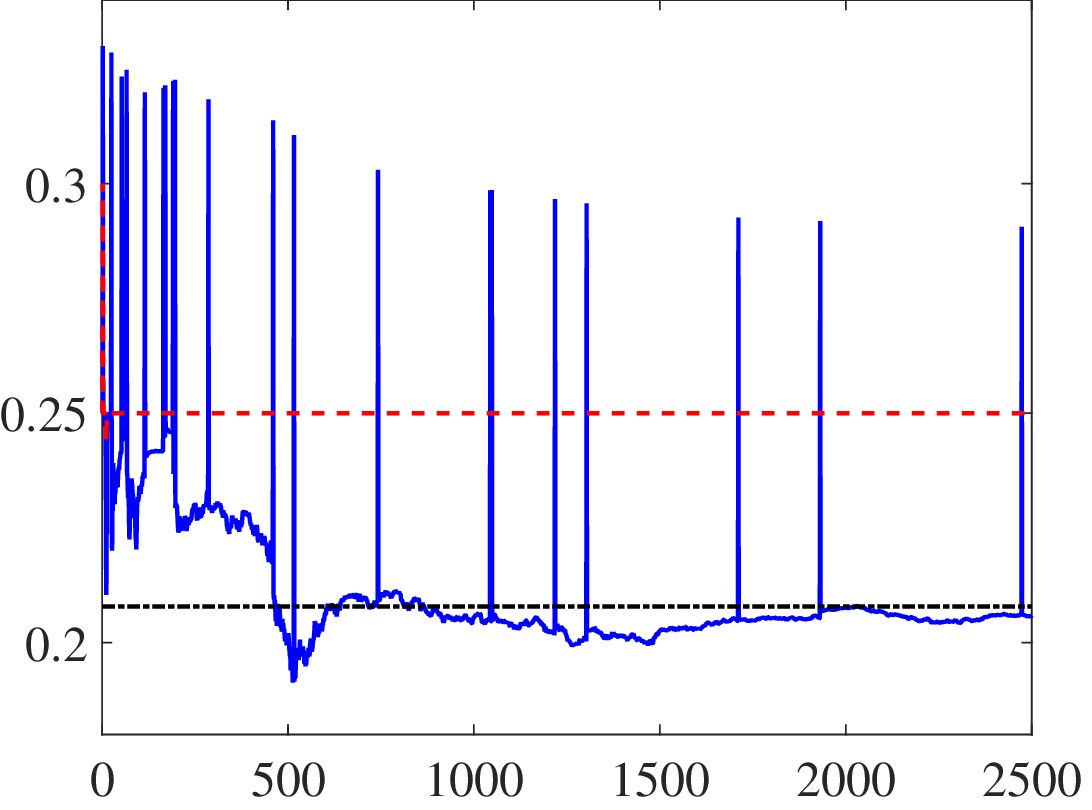}}}
\hspace{0.05em}
\subfloat{\label{fig:p}\fbox{\includegraphics[width=0.29\columnwidth]{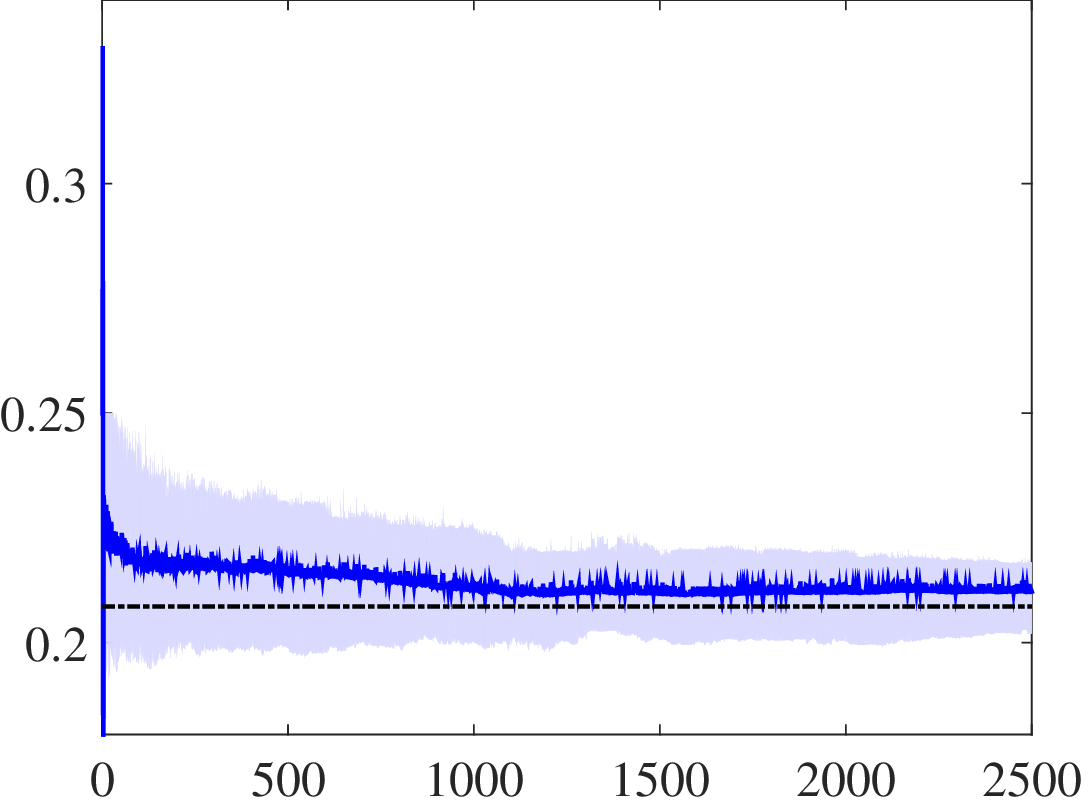}}}
\hspace{0.05em}
\subfloat{\label{fig:p_m}\fbox{\includegraphics[width=0.29\columnwidth]{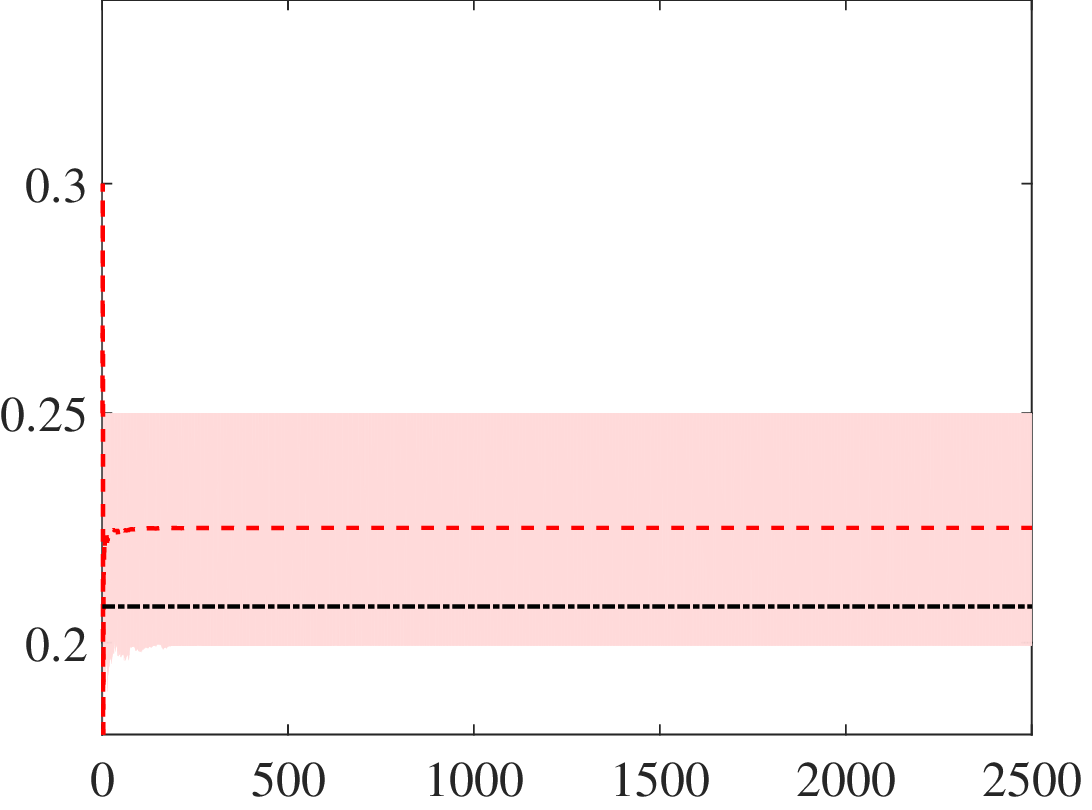}}} 

\vspace{0.5em}

\subfloat{\raisebox{0.11\columnwidth}[0pt][0pt]{ \makebox[0.059\columnwidth]{\small $\underset{\text{(kWh)}}{\quant_t}$}}} 
\hspace{0.02\columnwidth}
\subfloat{\label{fig:X_t}\fbox{\includegraphics[width=0.29\columnwidth]{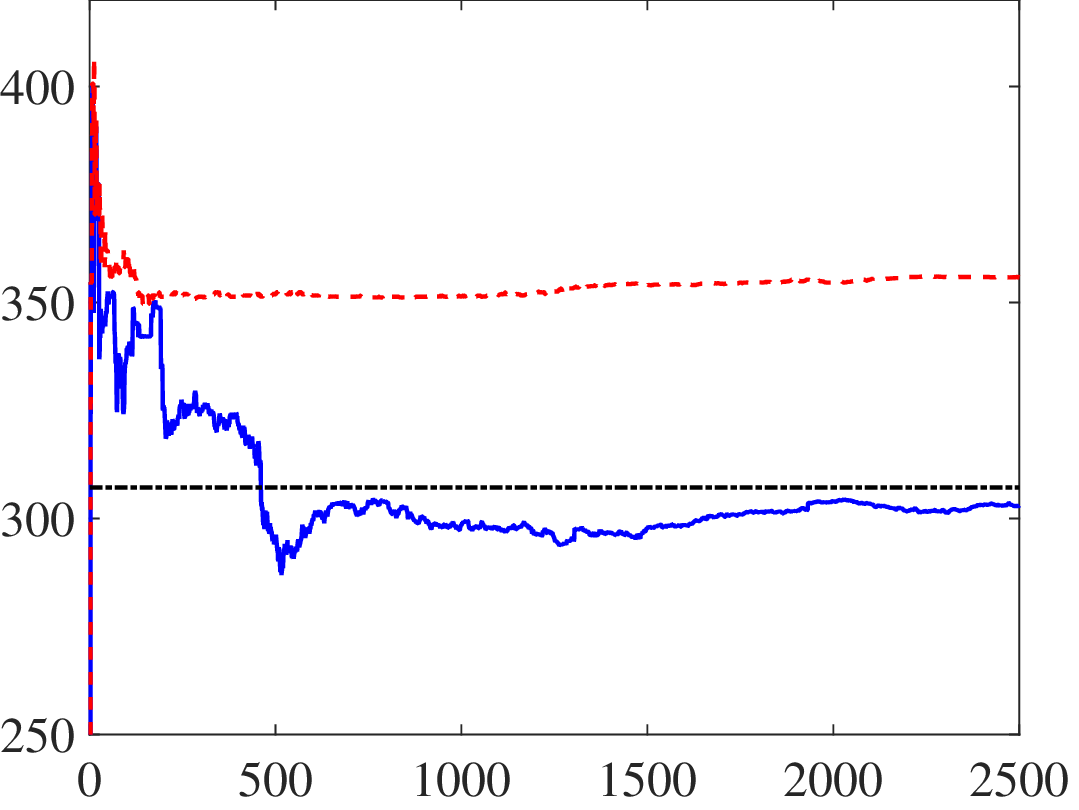}}}
\hspace{0.05em}
\subfloat{\label{fig:X}\fbox{\includegraphics[width=0.29\columnwidth]{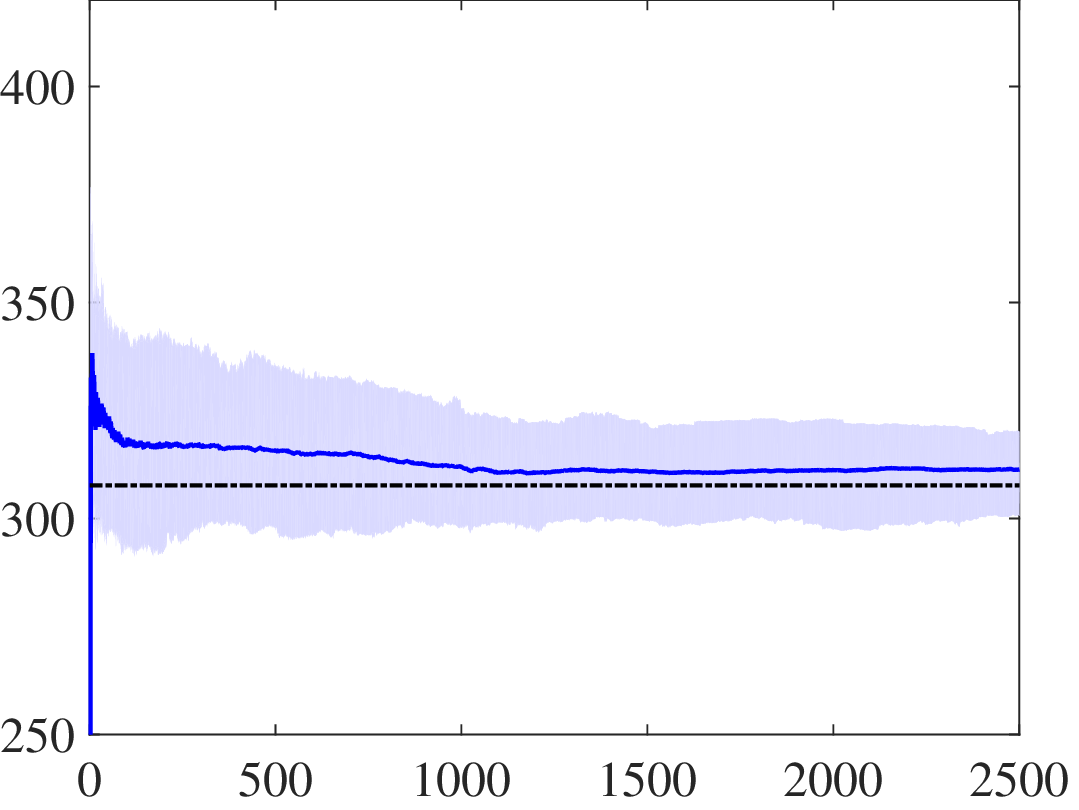}}}
\hspace{0.05em}
\subfloat{\label{fig:X_m}\fbox{\includegraphics[width=0.29\columnwidth]{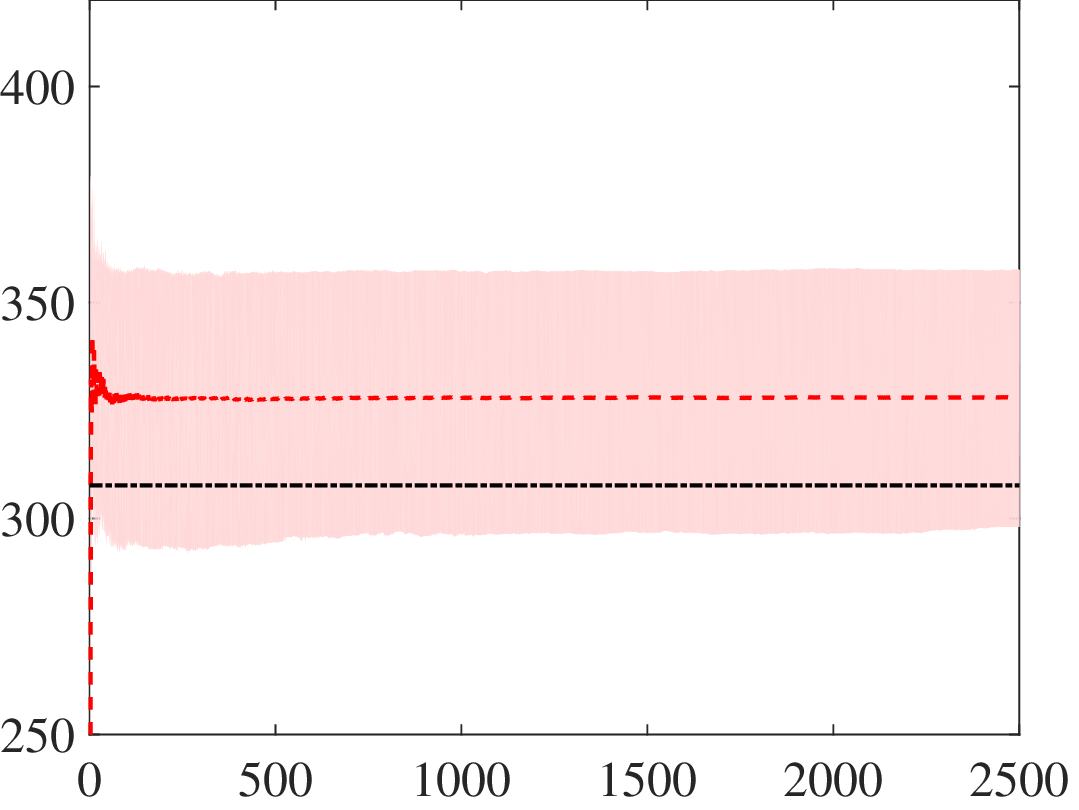}}} 

\vspace{0.5em}

\subfloat{\raisebox{0.11\columnwidth}[0pt][0pt]{ \makebox[0.059\columnwidth]{\small $\underset{\text{ (kWh$^2$/\$)}}{\widehat{a}_t}$}}} 
\hspace{0.02\columnwidth}
\subfloat{\label{fig:a_t}\fbox{\includegraphics[width=0.29\columnwidth]{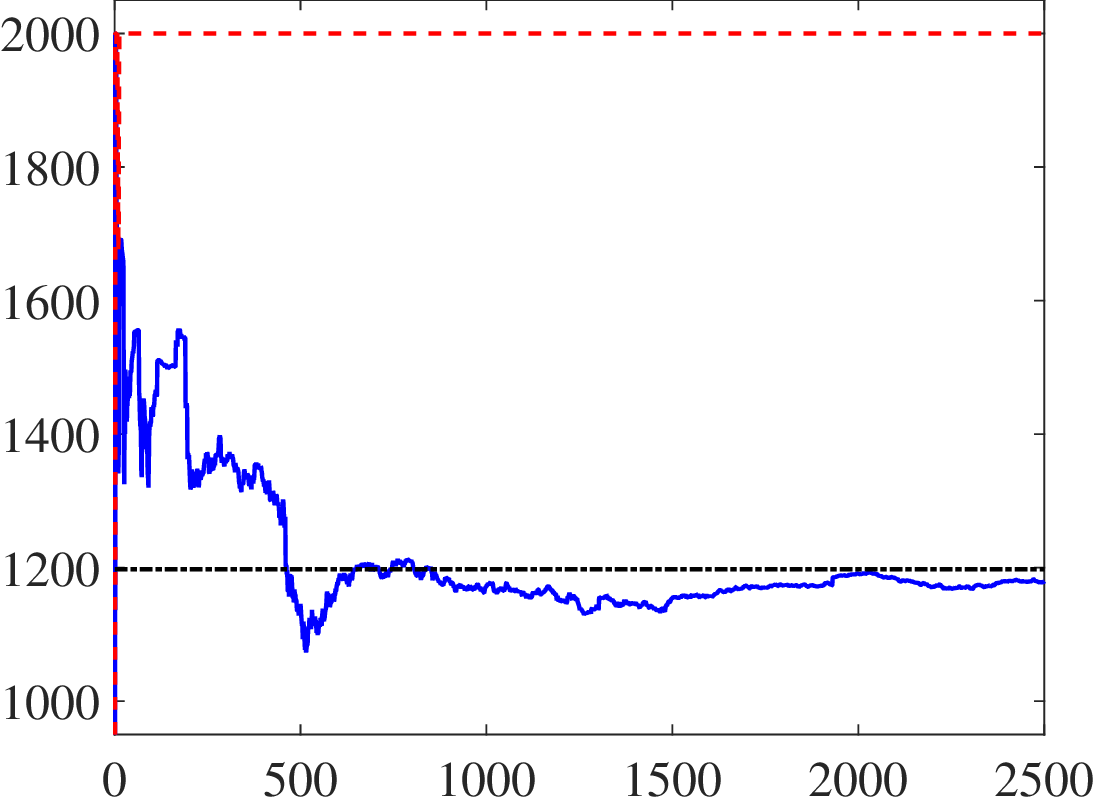}}} 
\hspace{0.05em}
\subfloat{\label{fig:a}\fbox{\includegraphics[width=0.29\columnwidth]{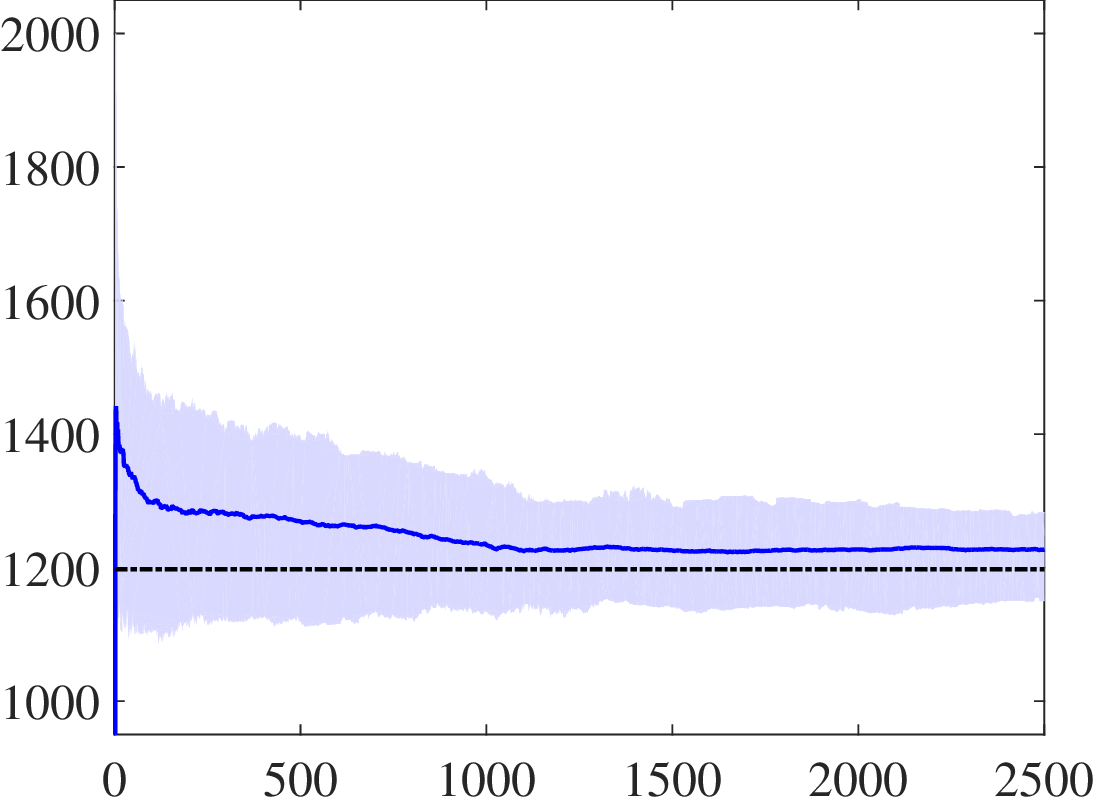}}} 
\hspace{0.05em}
\subfloat{\label{fig:a_m}\fbox{\includegraphics[width=0.29\columnwidth]{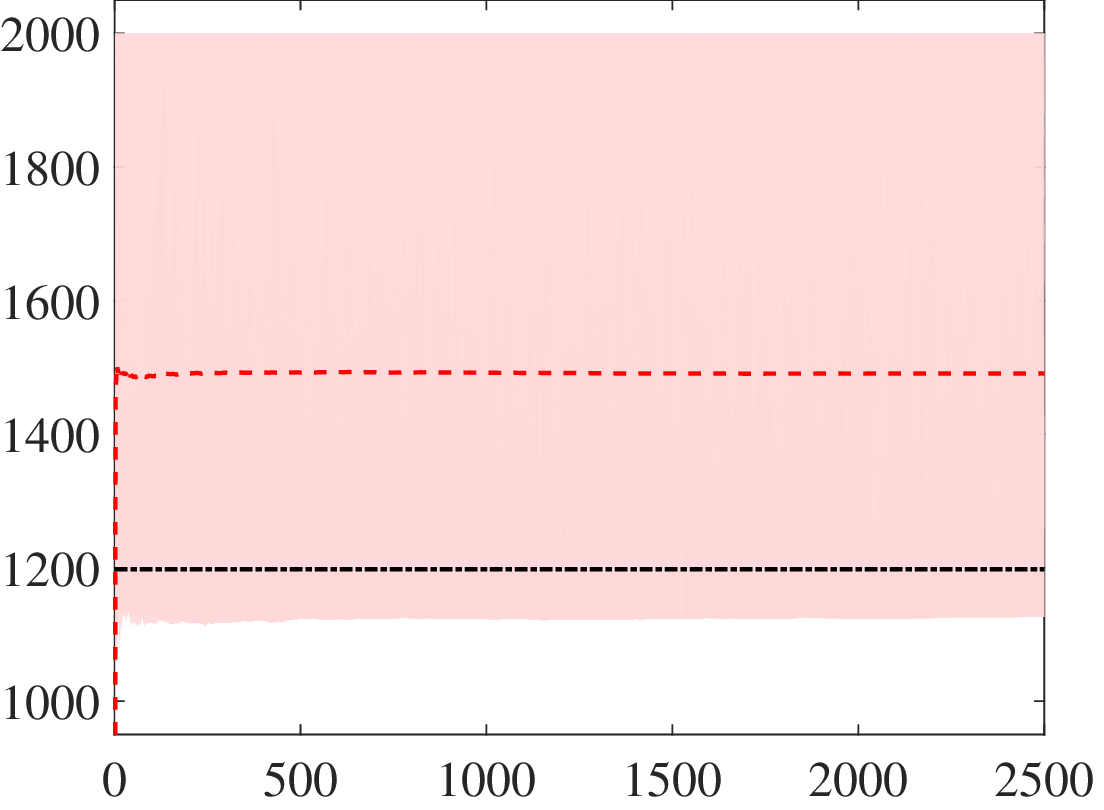}}}

\vspace{0.5em}

\subfloat{\raisebox{0.11\columnwidth}[0pt][0pt]{ \makebox[0.059\columnwidth]{\small  $\underset{\text{ (kWh)}}{\widehat{b}_t}$}}}  
\hspace{0.02\columnwidth}
\subfloat{\label{fig:b_t}\fbox{\includegraphics[width=0.29\columnwidth]{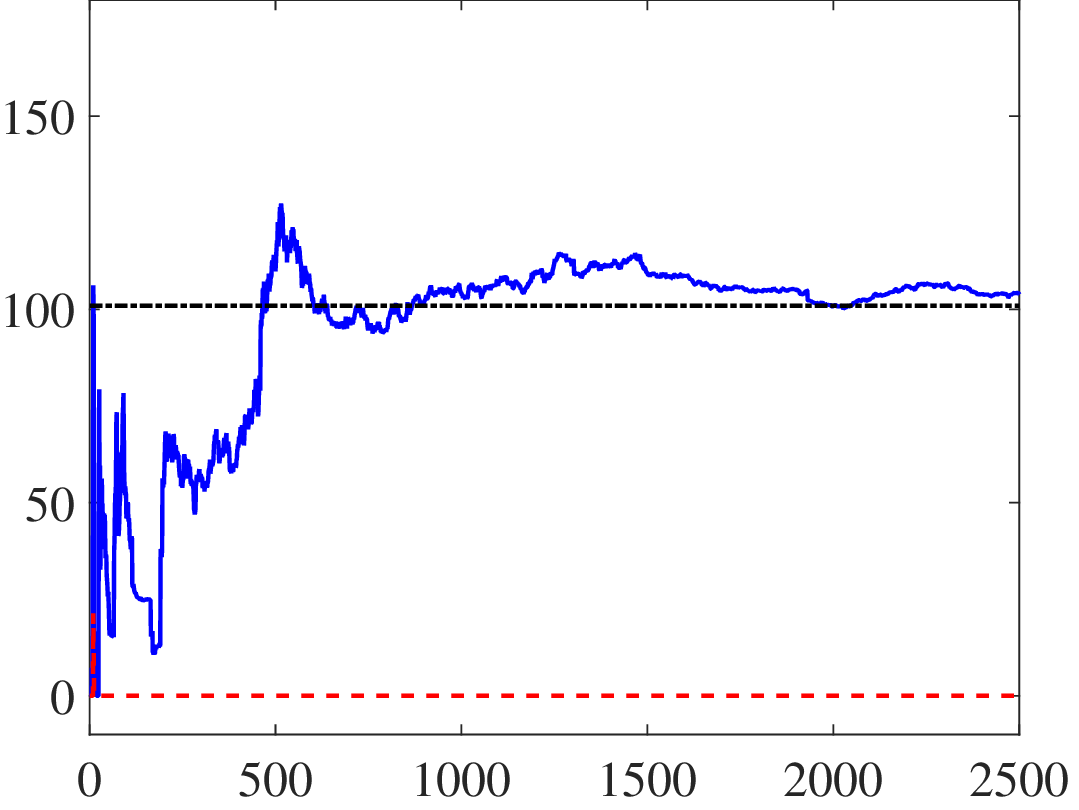}}}
\hspace{0.05em}
\subfloat{\label{fig:b}\fbox{\includegraphics[width=0.29\columnwidth]{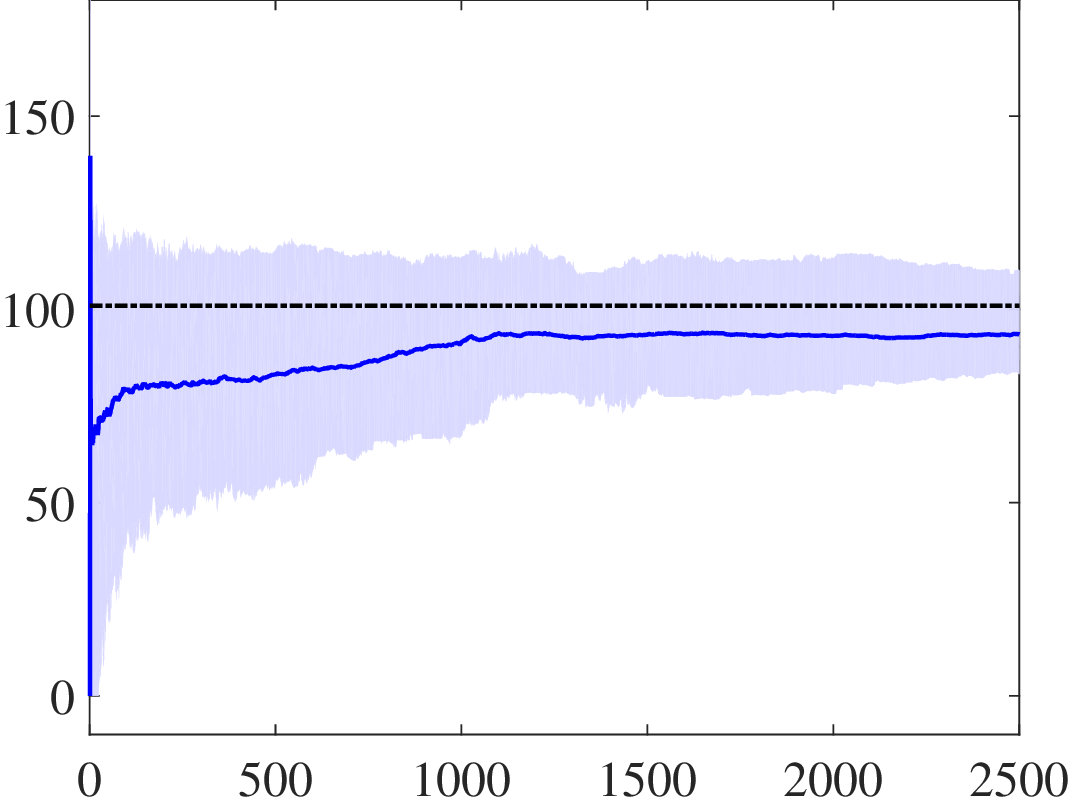}}}
\hspace{0.05em}
\subfloat{\label{fig:b_m}\fbox{\includegraphics[width=0.29\columnwidth]{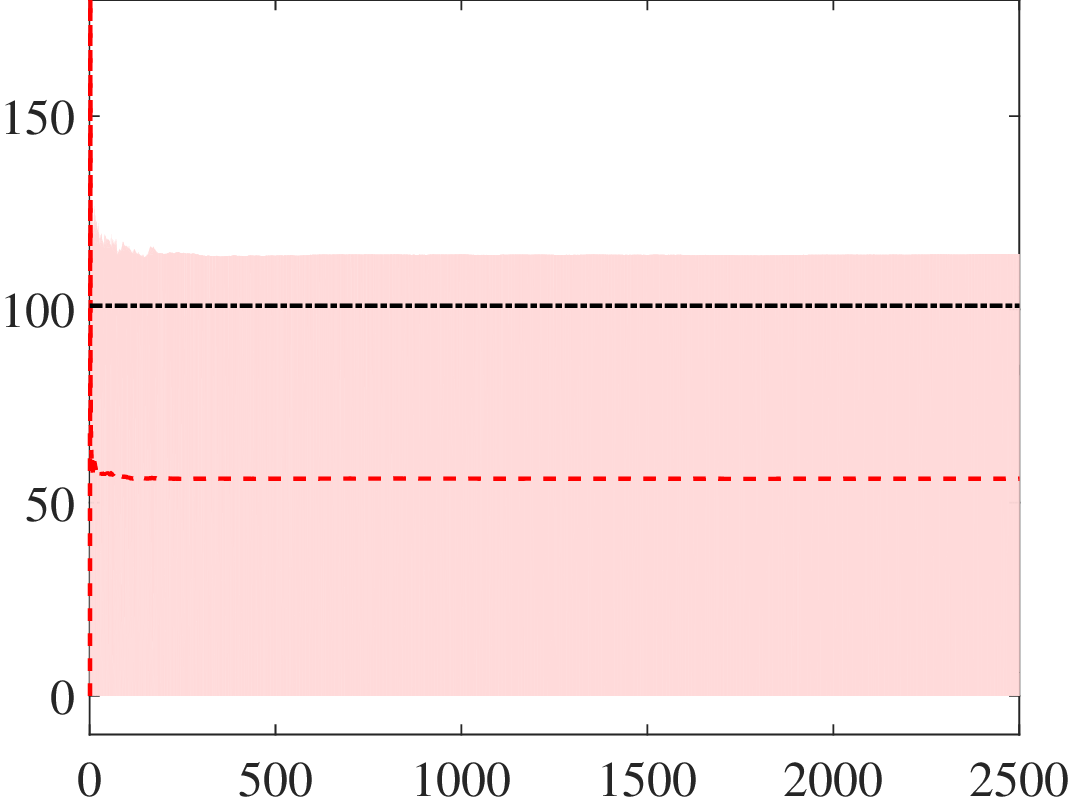}}}

\vspace{0.5em}

\subfloat{\raisebox{0.11\columnwidth}[0pt][0pt]{ \makebox[0.059\columnwidth]{\small  $\underset{\text{  (kWh)}}{\widehat{F}^{-1}_t(\alpha)}$}}}   
\hspace{0.02\columnwidth}
\subfloat[(a) Sample paths.]{\label{fig:Q_t}\fbox{\includegraphics[width=0.29\columnwidth]{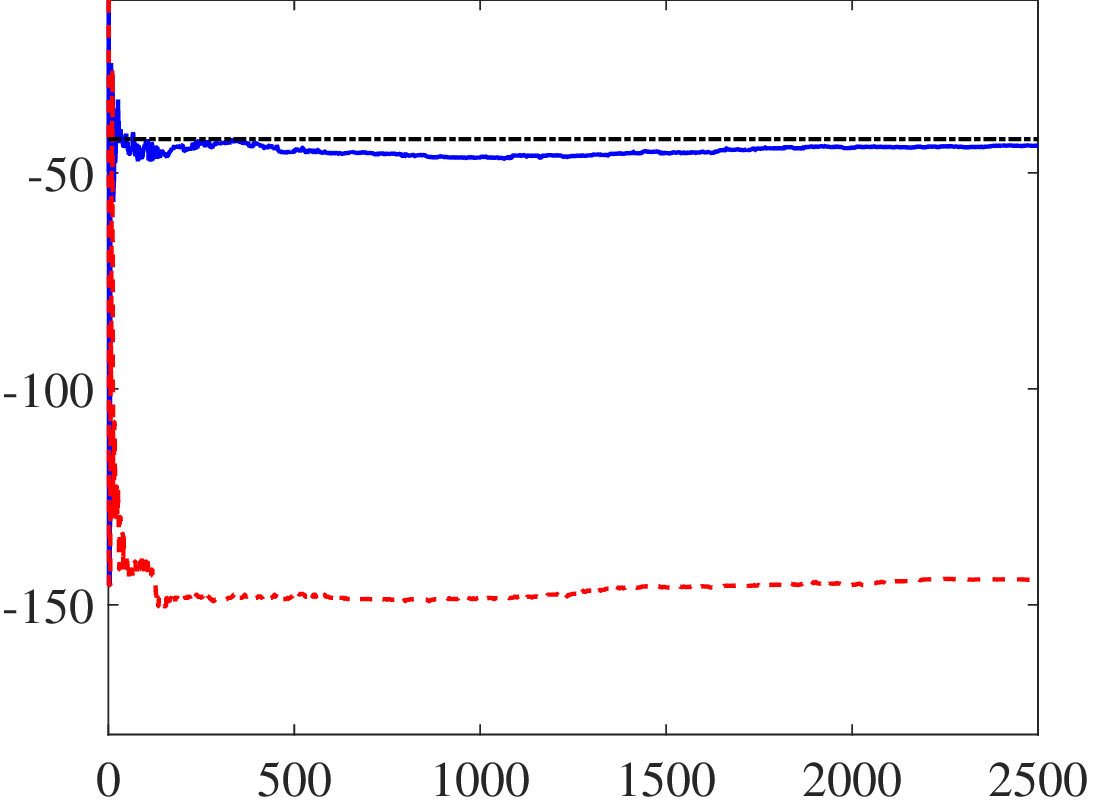}}}
\hspace{0.05em}
\subfloat[(b) RPMP.]{\label{fig:Q}\fbox{\includegraphics[width=0.29\columnwidth]{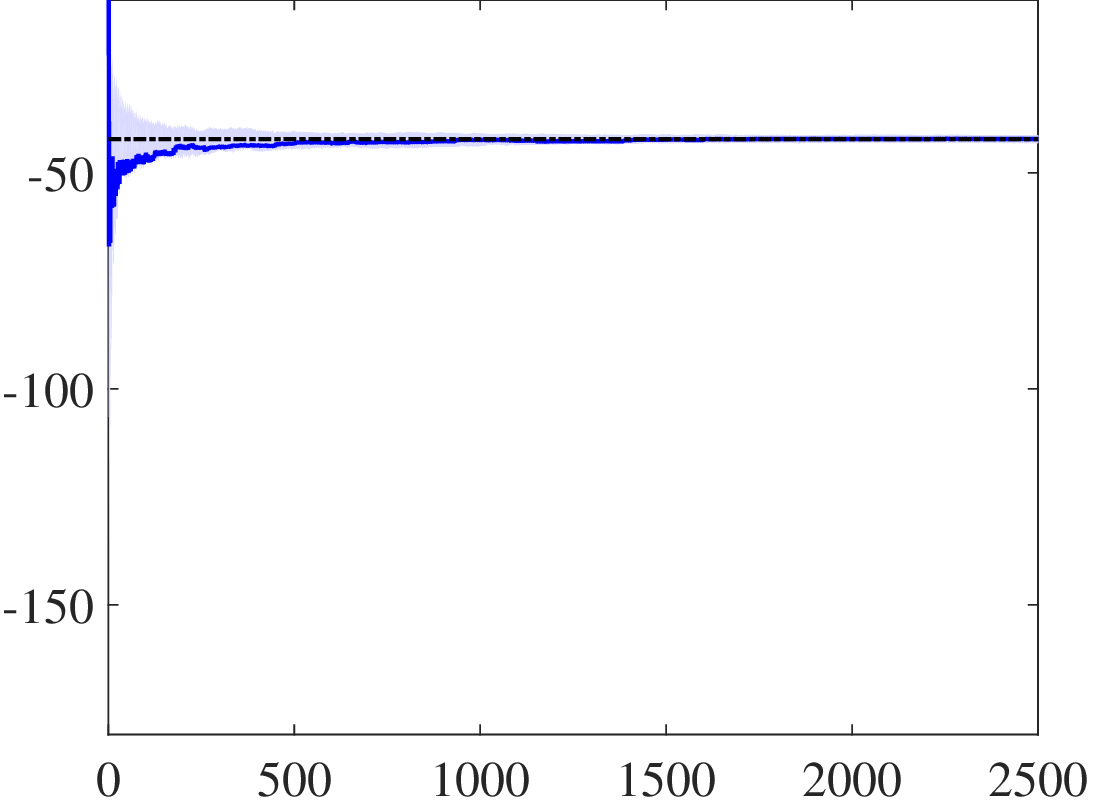}}}
\hspace{0.05em}
\subfloat[(c) MP.]{\label{fig:Q_m}\fbox{\includegraphics[width=0.29\columnwidth]{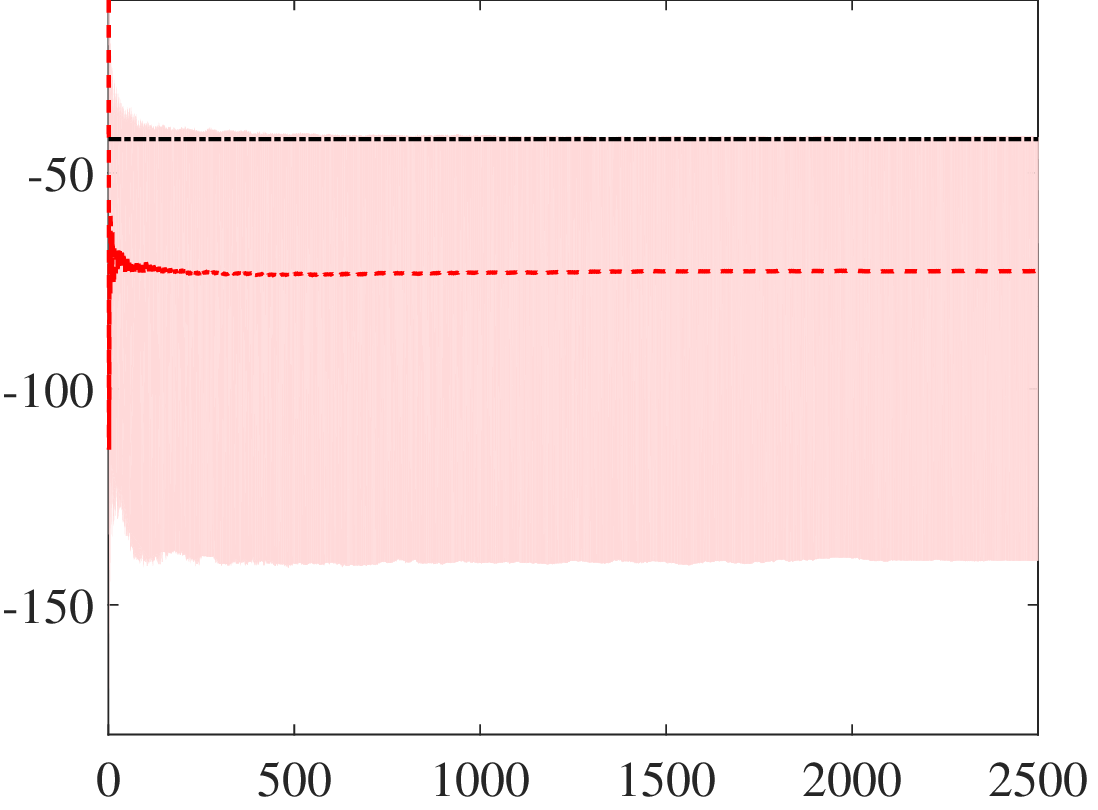}}}

\caption{Figures in the first column include plots of sample paths generated by the \emph{randomly perturbed myopic policy} (RPMP) (\blue{\hdashrule[0.5ex]{7mm}{1.25pt}{}}), the \emph{myopic policy} (MP) (\red{\hdashrule[0.5ex]{7mm}{1.25pt}{2.5pt}}), and the \emph{oracle optimal policy} (\hdashrule[0.5ex]{7mm}{1pt}{1pt}). Figures in the second and third columns include mean/confidence-interval plots associated with sequences generated by the RPMP (second column) and the MP (third column), compared against their oracle optimal policy counterparts. The shaded area represents their middle 70\% empirical confidence interval estimated using 100 independent experiments.}\label{fig:sequence}
\end{figure}


\subsection{Discussion}  
\begin{figure}
\centering
\includegraphics[width=0.5 \linewidth]{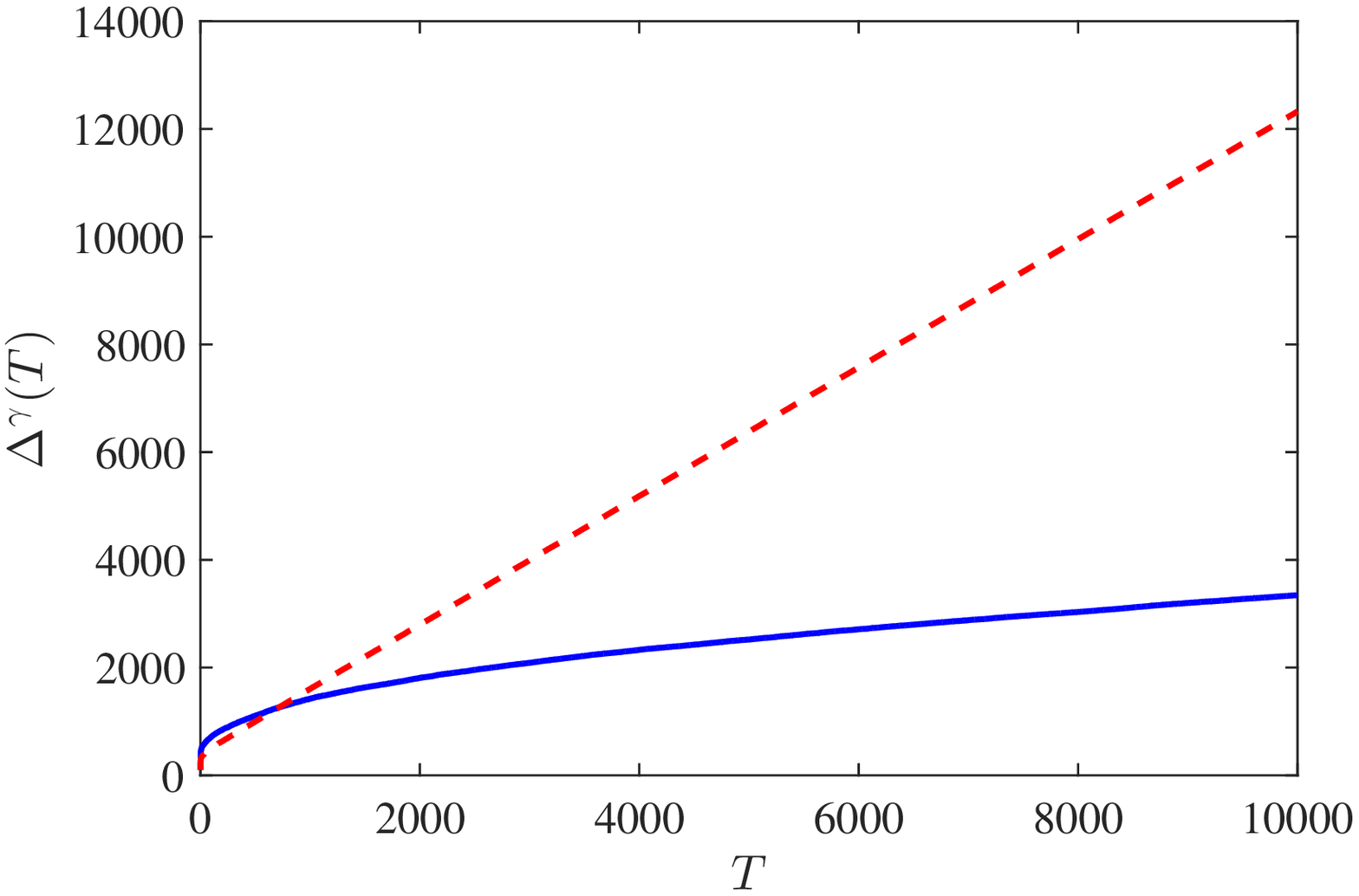}
\caption{A plot of the $T$-period regret incurred by the randomly perturbed myopic policy  (\blue{\hdashrule[0.5ex]{7mm}{1.25pt}{}}) compared to the $T$-period regret incurred by the myopic policy (\red{\hdashrule[0.5ex]{7mm}{1.25pt}{2.3pt}}).}\label{fig:regret}
\end{figure}

The plots in Figure \ref{fig:sequence}(c) illustrate an apparent lack of exploration in the sequence of posted prices generated by the myopic policy. That is to say, the myopic price sequence rapidly converges to a fixed value, which on average  differs substantially from the oracle optimal price. The same is true for the sequence of forward contracts generated by the myopic policy.
The premature convergence of the myopic price sequence, in turn, leads to incomplete learning with the parameter estimates converging incorrect values.  
As a consequence, the $T$-period regret incurred by the myopic policy grows linearly in $T$, as shown in Figure \ref{fig:regret}.

On the other hand, the persistent variation in the sequence of prices generated by the randomly perturbed myopic policy induces parameter estimates, which asymptotically converge to the true parameter values, which can be seen from the plots in Figure \ref{fig:sequence}(b). In particular, notice that the (middle 70\%) empirical confidence intervals   associated with the posted price and contract sequences generated by the randomly perturbed myopic policy shrink to their respective optimal oracle values over time. This provides empirical evidence supporting our theoretical claim that the sequences of prices and contracts generated by the randomly perturbed myopic policy converge to their oracle optimal values in the mean square sense.

\section{Conclusion}
In this paper, we study the problem of co-optimizing an aggregator's procurement and sale of demand response. The aggregator purchases energy in the form of demand reductions from a fixed group of residential customers, and sells the (a priori uncertain) aggregate demand reduction in a two-settlement wholesale electricity market. The customers' aggregate demand function is assumed to be affine in price (with unknown parameters) and subject to unobservable, additive random shocks (with unknown distribution). 
We propose a data-driven policy---referred to as the \emph{randomly perturbed myopic policy}---to guide the aggregator's adaptation of its posted DR prices and forward contract offerings over time. We show that the proposed policy  is consistent, meaning that the sequences of prices and contracts that it generates converge to the oracle optimal price and contract in the mean square sense. Moreover, we show that the regret incurred by the proposed policy over $T$ time periods is no more than $O(\log(T)\sqrt{T})$. 

As a direction for future research, it would be interesting to generalize the techniques developed in this paper to accommodate  time-varying and possibly nonlinear demand functions.

\Urlmuskip=0mu plus 1mu\relax
\bibliographystyle{elsarticle-num} 

\bibliography{references}{\markboth{References}{References}}                                                   

\begin{appendices}

\section{Notation}

\begin{table}[h]
\centering 
\begin{tabular}{l l}
\hline
\hline      \\ [-1.5ex]                 
Notation & Definition\vspace{1mm}\\  
\hline    \\ [-1.5ex]   
$p_t$ & DR price offered at time period $t$\\
$\quant_t$ & Forward contract commitment at time period $t$ \\        
$\pda_t$ & DA energy price at time period $t$\\
$\pover_t$ & RT overage price at time period $t$\\
$\pshort_t$ & RT shortage price at time period $t$\\
$\mu^+$& Expected value of the RT overage price at time period $t$\\
$\mu^-$& Expected value of the RT shortage price at time period $t$\\
$D_{t}$& Aggregate demand reduction at time period $t$\\
$\varepsilon_t$ & Aggregate demand shock at time period $t$\\
$F(\cdot)$ & Cumulative distribution function (CDF) of the aggregate \\ & demand  shock $\varepsilon_t$ at each time period $t$ \\
$\theta$& Demand model parameters $(a,b)$\\
$r_t(Q,p)$ & Expected profit of the aggregator at time period $t$\\
$\Delta^\gamma(T)$ & $T$-period regret incurred under feasible policy $\gamma$\vspace{1mm}\\ 
\hline   
\end{tabular}
\caption{Table of notations.}\label{tab:notation}
\end{table}

\section{Proof of Lemma \ref{lem:oracle}} \label{app:oracle}
Given a fixed pair $(Q,p)$, we have that
\begin{align*}
r_t(\quant,p)=&\pda_t Q -(ap^2+pb)+ \mu^+\mathbb{E}\left[ap+b-Q+\varepsilon_t\right]^+\\
&-\mu^- \mathbb{E}\left[Q-(ap+b)-\varepsilon_t\right]^+.
\end{align*}
It is straightforward to show that $r_t(\quant,p)$ is strictly concave in its arguments. It follows that one can characterize its unique maximizers as  solutions to the first order optimality conditions:
\begin{align}
\frac{\partial r_t(\quant,p)}{\partial p}&=-2ap-b+a\mu^+(1-F(\quant-ap-b))+a\mu^- F(\quant-ap-b)=0,  \label{eq:opt p}\\
\frac{\partial r_t(\quant,p)}{\partial \quant}&=\pda_t-\mu^+(1- F(\quant-ap-b))-\mu^- F(\quant-ap-b)=0. \label{eq:opt C}
\end{align}
The desired result follows.


\section{Proof of Lemma \ref{lem:delta}} \label{app:delta} 
Let $t \geq 1$, and fix $(Q_t,p_t)$. To streamline the proof, we define $Y_t := \quant_t - a p_t - b$ for each time period $t$. It follows that the expected profit of the aggregator can be expressed as
\begin{align*}
r_t (\quant_t,p_t) & =   \pda_t Y_t + \Embb\left[\mu^+ [\varepsilon_t - Y_t]^+ - \mu^- [Y_t - \varepsilon_t]^+ \right]+ (\pda_t - p_t) (a p_t + b).
\end{align*}
It will be helpful to decompose the expected profit as $r_t (\quant_t,p_t)  = r_{1t} (\quant_t,p_t) + r_{2t} (\quant_t,p_t)$, where
\begin{align*}
r_{1t} (\quant_t,p_t) &:= \pda_t Y_t +\Embb \left[  \mu^+ [\varepsilon_t - Y_t]^+ - \mu^- [Y_t - \varepsilon_t]^+ \right] ,\\
r_{2t} (\quant_t,p_t) &:= (\pda_t - p_t) (a p_t + b).
\end{align*}
It is straightforward to show that 
\begin{align}
r_{2t} (\quant^*_t,p^*_t) - r_{2t} (\quant_t,p_t) = a (p_t - p^*_t)^2. \label{eq:price_diff_profit}
\end{align}
Now we show that for each time period $t$, we have
\begin{align}
r_{1t} (\quant^*_t,p^*_t) - r_{1t} (\quant_t,p_t)\leq L(\mu^- - \mu^+) ( Y_t - Y^*_t )^2, \label{eq:ub_newsvendor_profit}
\end{align}
where $Y^*_t := \quant^*_t - ap^*_t - b$. It is straightforward to show that $Y^*_t= F^{-1} (\alpha_t)$. 
Consider first the case in which $Y_t \geq Y^*_t$. It follows that
\begin{align*}
&r_{1t} (\quant^*_t,p^*_t) - r_{1t} (\quant_t,p_t) \\
&=\pi_t(Y^*_t-Y_t)+\mu^+\int_{Y^*_t}^\infty (\varepsilon_t-Y^*_t)\ dF-\mu^+\int_{Y_t}^\infty (\varepsilon_t-Y_t)\ dF\\
&\quad -\mu^-\int_{-\infty}^{Y^*_t} (Y^*_t-\varepsilon_t)\ dF+\mu^-\int_{-\infty}^{Y_t} (Y_t-\varepsilon_t)\ dF \\
&= \pda_t (Y^*_t - Y_t) + \mu^+ \int_{Y^*_t}^{\infty} (Y_t - Y^*_t) \ dF + \mu^- \int_{-\infty}^{Y^*_t} (Y_t - Y^*_t)\ dF  \\
&\quad + (\mu^- - \mu^+)\int_{Y^*_t}^{Y_t}  (Y_t - \varepsilon_t)\ dF \\
&= (Y^*_t - Y_t) \big(\pda_t - \mu^+ (1 - F(Y^*_t)) - \mu^- F(Y^*_t) \big)  +(\mu^-- \mu^+)  \int_{Y^*_t}^{Y_t} (Y_t - \varepsilon_t)\ dF \\
&=  (\mu^- - \mu^+)\int_{Y^*_t}^{Y_t}  (Y_t - \varepsilon_t) \ dF,
\end{align*}
where the last equality follows from the fact that $F(Y^*_t) = F(F^{-1} (\alpha_t)) = \alpha_t$. Now, using the fact that $F$ is bi-Lipschitz, we obtain 
\begin{align*}
r_{1t} (\quant^*_t,p^*_t) - r_{1t} (\quant_t,p_t)   &= (\mu^- - \mu^+)\int_{Y^*_t}^{Y_t}  (Y_t - \varepsilon_t) \ dF\\
&\leq (\mu^- - \mu^+) (Y_t - Y^*_t) \int_{Y_t^*}^{Y_t} \ dF \\
&\leq L(\mu^- - \mu^+) (Y_t - Y^*_t)^2.
\end{align*}
For the case in which $Y_t < Y^*_t$, one can obtain an identical upper bound using  an analogous approach as above.
Finally, combining Inequality \eqref{eq:ub_newsvendor_profit} and Equation \eqref{eq:price_diff_profit} with the fact that $Y_t-Y^*_t=Q_t-Q^*_t-a(p_t-p^*_t)$ yields the desired upper bound on regret.


\section{Proof of Lemma \ref{lem:pvariance}} \label{app:pvariance}
To simplify the presentation in the sequel, we define $u_t$ and $U_t$ as follows
\begin{align*}
u_t:=\sum_{k=1}^{t} \bmat{p_k \\ 1} \varepsilon_k\quad \text{and}\quad U_t:=u_t^\top P_t^{-1}u_t.
\end{align*}
Recall from Equation \eqref{eq:LSE} that $\theta_t-\theta=P_t^{-1}u_t$. Then,
\begin{align*}
\mathbb{E}^\gamma\left[\|\widehat{\theta}_t-\theta\|^2\right]&\leq \mathbb{E}^\gamma\left[\|P_{t}^{-1} u_t\|^2\right]\leq  \mathbb{E}^\gamma\left[\|P_{t}^{-1/2}\|^2\|P_{t}^{-1/2} u_t\|^2\right],
\end{align*}
where the last inequality follows from the Cauchy-Schwarz inequality. Using the definition of matrix norms, it holds that
\begin{align*}
\|P_{t}^{-1/2}\|^2=\left(\lambda_\mathsf{max}\left(P_t^{-1/2}\right)\right)^2=\lambda_\mathsf{min}(P_t)^{-1},
\end{align*} 
where $\lambda_\mathsf{max}(A)$ and $\lambda_\mathsf{min}(A)$ are defined as the largest and smallest eigenvalues of the matrix $A$, respectively. We construct a lower bound on $\lambda_\mathsf{min}(P_t)$ similar to \citep[Lemma 1]{den2013simultaneously}. To bound $\lambda_\mathsf{min}(P_t)$, we first find the characteristic polynomial of $P_t$. Recall the definition of $P_t$, that is
\begin{align*}
P_t=\sum_{k=1}^{t} \bmat{p_k \\ 1} \bmat{p_k \\ 1}^\top = \bmat{\sum_{k=1}^t p_k^2 & \sum_{k=1}^t p_k \\\sum_{k=1}^t p_k & t } = \bmat{\sum_{k=1}^t p_k^2 & t\bar{p}_t \\t\bar{p}_t & t }.
\end{align*}
The characteristic polynomial of $P_t$ is given by
\begin{align}
\lambda^2-\lambda \left(t+\sum_{k=1}^t p_k^2\right)+t^2V_t=0. \label{eq:char poly}
\end{align}
From Equation \eqref{eq:char poly} it follows that
\begin{align}
\lambda_\mathsf{max}(P_t) + \lambda_\mathsf{min}(P_t) &= t+\sum_{k=1}^t p_k^2,\label{eq:sum eig}\\
\lambda_\mathsf{max}(P_t) \lambda_\mathsf{min}(P_t) &= t^2V_t. \label{eq:mul eig}
\end{align}
Define $\overline{p}$ as $\overline{p}:=\mu^-/2+\rho$, which upper bounds prices generated under the RPMP policy. From Equation \ref{eq:sum eig} it follows that $\lambda_\mathsf{max}(P_t)\leq (1+\overline{p}^2)t$. Thus, using Equation \ref{eq:mul eig}, we get
\begin{align*}
\lambda_\mathsf{min}(P_t)\geq \frac{tV_t}{1+\overline{p}^2}.
\end{align*}
Using the lower bound on the minimum eigenvalue of $P_t$ and the definition of $U_t$, we get
\begin{align}
\mathbb{E}^\gamma\left[\|\widehat{\theta}_t-\theta\|^2\right]&\leq \mathbb{E}^\gamma\left[\lambda_\mathsf{min}(P_t)^{-1} \|P_{t}^{-1/2} u_t\|^2\right]\nonumber \\
&\leq (1+\overline{p}^2)\mathbb{E}^\gamma\left[\frac{U_t}{tV_t}\right]. \label{eq:bound mse q}
\end{align}
Now we establish a result that lower bounds the price variance $V_t$ by a function of random variables $\xi_1,\ldots,\xi_t$. Its proof is postponed to Appendix \ref{app:V Xi}.
\begin{lemm}[Bound on $V_t$]\label{lem:V Xi}
 Under the randomly perturbed myopic policy \eqref{pol:perturbed q_1} and \eqref{pol:perturbed p_1}, the price variance $V_t$ is lower bounded by
\begin{align}
V_t\geq \frac{\Xi_t}{t}\quad \text{almost surely}, \label{eq:bound xi}
\end{align}
for $t\geq 3$. Here, $\Xi_t$ is defined as 
\begin{align}
\Xi_t := \frac{1}{2}(p_2-p_1)^2 +\frac{2}{3}\rho^2 \sum_{k=3}^t\xi_k. \label{eq:Xi}
\end{align}
\end{lemm}
By applying Inequality \eqref{eq:bound xi} to Inequality \eqref{eq:bound mse q}, we get
\begin{align}
\mathbb{E}^\gamma\left[\|\widehat{\theta}_t-\theta\|^2\right]\leq (1+\overline{p}^2)\mathbb{E}^\gamma\left[\frac{U_t}{\Xi_t}\right]. \label{eq:bound mse xi}
\end{align}
We now bound $\mathbb{E}^\gamma[U_t/\Xi_t]$ similar to Lai and Wei \cite[Theorem 1]{lai1982least}. They use the extended stochastic Liapounov functions and construct a recursive bound on $U_t$. Using a similar argument we construct a recursive bound on $\mathbb{E}^\gamma[U_k/\Xi_t]$ for $k=3,\ldots,t$. The proof of Lemma \ref{lem:laiwei} is postponed to Appendix \ref{app:laiwei}. 
\begin{lemm}\label{lem:laiwei}
For all $t\geq 3$, it holds that
\begin{align}
\mathbb{E}^\gamma\left[\frac{U_t}{\Xi_t}\right]\leq 2\sigma^2\expe{\frac{1}{\Xi_t}} + \sigma^2\sum_{k=3}^t  \mathbb{E}^\gamma\left[\frac{1}{\Xi_t}\bmat{p_k \\ 1}^\top P_k^{-1}\bmat{p_k \\ 1}\right], \label{eq:bound U}
\end{align}
where $\sigma^2:=\mbox{Var}(\varepsilon_t)\leq (\overline{\varepsilon}-\underline{\varepsilon})^2$.
\end{lemm}
Now we upper bound the second term in the right hand side of Inequality \eqref{eq:bound U}. It is straightforward to show that 
\begin{align*}
\bmat{p_k \\ 1}^\top P_k^{-1}\bmat{p_k \\ 1}&= \frac{1}{kV_k}\bmat{p_k \\ 1}^\top \bmat{1&-\bar{p}_k \\ -\bar{p}_k  &V_k+\bar{p}_k^2}\bmat{p_k \\ 1}=1-\frac{(k-1)V_{k-1}}{kV_k}+\frac{1}{k}.
\end{align*}
Note that for all $x>0$, it holds that $1-x\leq \log(1/x)$. Thus,
\begin{align}
\sum_{k=3}^t\bmat{p_k \\ 1}^\top P_k^{-1}\bmat{p_k \\ 1}&\leq \sum_{k=3}^t \left(\log\left(\frac{kV_k}{(k-1)V_{k-1}}\right)+\frac{1}{k}\right)\nonumber\\
&=  \log\left(\frac{tV_t}{2V_2}\right)+\sum_{k=3}^t \frac{1}{k}\nonumber\\
&\leq \log\left(\frac{tV_t}{2V_2}\right) + \log(t)\nonumber\\
&\leq \log\left(\frac{2\overline{p}^2}{(p_2-p_1)^2}\right)+2\log(t) \quad \text{almost surely},\label{eq:bound log p}
\end{align}
where the last inequality follows from the fact that $V_t\leq \overline{p}^2$ almost surely and $V_2=(1/4)(p_2-p_1)^2$.
By applying Inequality \eqref{eq:bound log p} and  \eqref{eq:bound U} to Inequality \eqref{eq:bound mse xi}, we get
\begin{align}
\mathbb{E}^\gamma&\left[\|\widehat{\theta}_t-\theta\|^2\right]\nonumber \\
&\leq  2\sigma^2(1+\overline{p}^2)\left(1+\log\left(\frac{\sqrt{2}\overline{p}}{|p_2-p_1|}\right)+\log(t)\right) \mathbb{E}^\gamma\left[\frac{1}{\Xi_t}\right]\nonumber\\
&\leq 2\sigma^2(1+\overline{p}^2)\left( 1 +  \frac{1}{\log(3)}\log\left(\frac{2\sqrt{2}\overline{p}}{|p_2-p_1|}\right) \right) \log(t)\mathbb{E}^\gamma\left[\frac{1}{\Xi_t}\right], \label{eq:bound 1/xi}
\end{align}
where the last inequality follows from the fact that $\log(t)+c\leq (1+c/\log(3))\log(t)$ for $t\geq 3$ and all $c\geq 0$, and the fact that $|p_2-p_1|\leq \overline{p}$. Finally, we establish an upper bound on $\mathbb{E}^\gamma[{1}/{\Xi_t}]$ in Lemma \ref{lem:1/xi}, which its proof is given in Appendix \ref{app:1/xi}.
\begin{lemm}\label{lem:1/xi}
For $t\geq 3$, it holds that 
\begin{align*}
\expe{\frac{1}{\Xi_t}} \leq \frac{3}{(p_2-p_1)^2\eta t^{1-r}}.
\end{align*}
\end{lemm}
Setting as follows yields the desired result.
\begin{align*}
K:=\frac{6\sigma^2(1+\overline{p}^2)}{(p_2-p_1)^2\eta}\left( 1 +  \frac{1}{\log(3)}\log\left(\frac{2\sqrt{2}\overline{p}}{|p_2-p_1|}\right) \right).
\end{align*}


\section{Proof of Theorem \ref{thm:regret}} \label{app:regret}
We first establish a result that relates pricing and contracting errors under the RPMP to the parameter estimation error. Its proof is postponed to Appendix \ref{app:errors}.

\begin{lemm}\label{lem:errors}
Under the randomly perturbed myopic policy \eqref{pol:perturbed q_1} and \eqref{pol:perturbed p_1}, it holds that
\begin{align}
\mathbb{E}^\gamma\left[(p_t-p^*_t)^2\right]\leq k_1\mathbb{E}^\gamma\left[\|\widehat{\theta}_{t-1}-\theta\|^2\right]+ k_2 t^{-r}, \label{eq:first bound}
\end{align}
and
\begin{align}
&\mathbb{E}^\gamma\left[(\quant_t-\quant^*_t-a(p_t-p^*_t))^2\right]\\
&\qquad\leq k_3\mathbb{E}^\gamma\left[\| \widehat{\theta}_{t-1}-\theta\|^2\right] + k_4 \frac{1}{t-1}  + k_5  \frac{1}{\sqrt{t-1}}+k_6 t^{-r}, \label{eq:second bound}
\end{align}
for all $t\geq 3$.
\end{lemm}

We combine  Lemmas \ref{lem:delta} and  \ref{lem:errors} to obtain
\begin{align}
\Delta^\gamma(T)\leq&  k_0 +  (k_1+k_3L(\mu^- - \mu^+)) \sum_{t=3}^{T} \mathbb{E}^\gamma\left[\|\widehat{\theta}_{t-1}-\theta\|_1^2\right]\nonumber\\
& +\sum_{t=3}^T  (k_2+k_6 L(\mu^- - \mu^+))t^{-r} + L(\mu^- - \mu^+)\left(k_4\frac{1}{t-1}+k_5  \frac{1}{\sqrt{t-1}}\right)\nonumber\\
\leq & k_0 +  (k_1+k_3L(\mu^- - \mu^+)) \sum_{t=2}^{T} \mathbb{E}^\gamma\left[\|\widehat{\theta}_{t}-\theta\|_1^2\right]\nonumber\\
& +(k_2+k_6 L(\mu^- - \mu^+))\sum_{t=3}^T t^{-r} + L(\mu^- - \mu^+)\sum_{t=2}^T\left(k_4\frac{1}{t}+k_5  \frac{1}{\sqrt{t}}\right)\label{eq:delta primary},
\end{align}
where $k_0 := \sum_{t=1}^2 a(p_t-p^*_t)^2 + L(\mu^- - \mu^+)(\quant_t-\quant^*_t-a(p_t-p^*)^2)$. Now by applying the bound on the parameter estimation error in Lemma \ref{lem:mse} to Inequality \eqref{eq:delta primary}, we get
\begin{align*}
\Delta^\gamma(T)\leq & k_0 + (k_1+k_3L(\mu^- - \mu^+))K \sum_{t=2}^{T} \frac{\log(t)}{t^{1-r}}\\
& + (k_2+k_6 L(\mu^- - \mu^+))\sum_{t=3}^{T}t^{-r} +L(\mu^- - \mu^+)\sum_{t=2}^T   \left(k_4\frac{1}{t}+k_5  \frac{1}{\sqrt{t}}\right) \\
\leq & k_0 + (k_1+k_3L(\mu^- - \mu^+))K\frac{\log(T)}{r}T^r+ \frac{k_2+k_6 L(\mu^- - \mu^+)}{1-r}T^{1-r} \\
& + L(\mu^- - \mu^+)\left(k_4\log(T)+2k_5  \sqrt{T}\right) \\
\leq & C_0 + C_1\log(T) + \left(\frac{C_2}{1-r}\right)T^{1-r} +  \left(\frac{C_2}{r}\right)\log(T)T^r.
\end{align*}
The second inequality follows from the fact that for $r\in(0,1)$, we have that
\begin{align*}
\sum_{t=2}^T \frac{\log(t)}{t^{1-r}} \leq \int_1^T \frac{\log(x)}{x^{1-r}} dx\leq  \frac{1}{r} \log(T)T^r, 
\end{align*}
and
\begin{align*}
\sum_{t=3}^T t^{-r} \leq \int_2^T x^{-r}dx \leq  \frac{1}{1-r} T^{1-r},
\end{align*}
and that
\begin{align*}
\sum_{t=2}^T  \frac{1}{t} \leq \int_1^T \frac{1}{x}dx =   \log(T).
\end{align*}
We complete the proof by defining the constants $C_0$, $C_1$, and $C_2$ as follows.
\begin{align*}
C_0&:=k_0,\\
C_1&:=L(\mu^- - \mu^+)k_4,\\
C_2&:=2L(\mu^- - \mu^+)k_5 + \max\left\{k_2+k_6 L(\mu^- - \mu^+),(k_1+k_3L(\mu^- - \mu^+))K\right\}.
\end{align*}


\section{Proof of Lemma \ref{lem:V Xi}} \label{app:V Xi}

It is straightforward to show that
\begin{align*}
\sum_{k=1}^t(p_k-\bar{p}_t)^2=\sum_{k=2}^t\frac{k-1}{k}(p_k-\bar{p}_{k-1})^2.
\end{align*}
Then for the RPMP, we get
\begin{align*}
tV_t&=\sum_{k=2}^t\frac{k-1}{k}(p_k-\bar{p}_{k-1})^2\\
&=\frac{1}{2}(p_2-p_1)^2 + \sum_{k=3}^t\frac{k-1}{k}(p_k-\bar{p}_{k-1})^2\\
&=\frac{1}{2}(p_2-p_1)^2 + \sum_{k=3}^t\frac{k-1}{k}(\rho \xi_k + (\widehat{p}_k-\bar{p}_{k-1})(1-\xi_k))^2\\
&=\frac{1}{2}(p_2-p_1)^2 + \sum_{k=3}^t\frac{k-1}{k}\rho^2 \xi_k^2 + \sum_{k=3}^t\frac{k-1}{k}(\widehat{p}_k-\bar{p}_{k-1})^2(1-\xi_k)^2,
\end{align*}
where the last equality follows from the fact that $\xi_k(1-\xi_k)=0$ for all $k\geq 3$. Thus, almost surely it holds that
\begin{align*}
tV_t &\geq \frac{1}{2}(p_2-p_1)^2 +\sum_{k=3}^t\frac{k-1}{k}\rho^2 \xi_k^2\\
     &\geq \frac{1}{2}(p_2-p_1)^2 +\frac{2}{3}\rho^2 \sum_{k=3}^t\xi_k^2\\
     &=    \frac{1}{2}(p_2-p_1)^2 +\frac{2}{3}\rho^2 \sum_{k=3}^t\xi_k,
\end{align*}
where the last equality follows from the fact that $\xi_k$ only takes values in $\{0,1\}$. 


\section{Proof of Lemma \ref{lem:laiwei}} \label{app:laiwei}
Recall from the definition of $\Xi_t$ in Equation \eqref{eq:Xi} that it is only a function of $\xi_3,\ldots,\xi_t$. Thus $\varepsilon_k$ is independent of $\Xi_t$ for all $k=1,\ldots,t$. Then,
\begin{align}
\expe{\frac{U_k}{\Xi_t}}&=\expe{\frac{1}{\Xi_t}\left(u_{k-1}+\bmat{p_k \\ 1} \varepsilon_k\right)^\top P_k^{-1}\left(u_{k-1}+\bmat{p_k \\ 1} \varepsilon_k\right)}\nonumber\\
&=\expe{\frac{1}{\Xi_t}u_{k-1}^\top P_k^{-1}u_{k-1}} + 2 \expe{\frac{1}{\Xi_t}u_{k-1}^\top P_k^{-1}\bmat{p_k \\ 1} \varepsilon_k}\nonumber\\
&\quad+\expe{\frac{1}{\Xi_t}\bmat{p_k \\ 1}^\top P_k^{-1}\bmat{p_k \\ 1} \varepsilon_k^2} \nonumber\\
&=\expe{\frac{1}{\Xi_t}u_{k-1}^\top P_k^{-1}u_{k-1}}\nonumber \\
&\quad+ 2 \expe{\frac{1}{\Xi_t}u_{k-1}^\top P_k^{-1}\bmat{p_k \\ 1} \expe{\varepsilon_k\ \Big|\ \Xi_t,p_1,\ldots,p_k,\varepsilon_1,\ldots,\varepsilon_{k-1}}}\nonumber\\
&\quad+\expe{\frac{1}{\Xi_t}\bmat{p_k \\ 1}^\top P_k^{-1}\bmat{p_k \\ 1} \expe{\varepsilon_k^2\ \Big|\  \Xi_t,p_1,\ldots,p_k,\varepsilon_1,\ldots,\varepsilon_{k-1}}}\nonumber\\
&=\expe{\frac{1}{\Xi_t}u_{k-1}^\top P_k^{-1}u_{k-1}}+\sigma^2\expe{\frac{1}{\Xi_t}\bmat{p_k \\ 1}^\top P_k^{-1}\bmat{p_k \\ 1}}\label{eq:U Xi},
\end{align}
where the third equality is a direct application of the law of iterated expectations. The last equality follows from the fact that $\expe{\varepsilon_k|  \Xi_t,p_1,\ldots,p_k,\varepsilon_1,\ldots,\varepsilon_{k-1}}=0$ and $\expe{\varepsilon_k^2|\Xi_t,p_1,\ldots,p_k,\varepsilon_1,\ldots,\varepsilon_{k-1}}=\mbox{Var}(\varepsilon_k)=\sigma^2$. 
 Using Sherman-Morrison formula we get
\begin{align*}
P_k^{-1}&=\left(P_{k-1}+\bmat{p_k \\ 1}\bmat{p_k \\ 1}^\top\right)^{-1}=P_{k-1}^{-1} - \frac{P_{k-1}^{-1}\bmat{p_k \\ 1}\bmat{p_k \\ 1}^\top  P_{k-1}^{-1}}{1+\bmat{p_k \\ 1}^\top P_{k-1}^{-1}\bmat{p_k \\ 1}}.
\end{align*}
It follows that,
\begin{align*}
\expe{\frac{1}{\Xi_t}u_{k-1}^\top P_k^{-1}u_{k-1}} &= \expe{\frac{U_{k-1}}{\Xi_t}} - \expe{\frac{1}{\Xi_t}\frac{\left(u_{k-1}P_{k-1}^{-1}\bmat{p_k \\ 1}\right)^2}{1+\bmat{p_k \\ 1}^\top P_{k-1}^{-1}\bmat{p_k \\ 1}}}\\
&\leq  \expe{\frac{U_{k-1}}{\Xi_t}},
\end{align*}
where the inequality follows from the fact that the random variable in the second expectation is  non-negative almost surely. By applying the above inequality to Inequality \eqref{eq:U Xi}, we get 
\begin{align*}
\expe{\frac{U_k}{\Xi_t}}-\expe{\frac{U_{k-1}}{\Xi_t}}\leq  \sigma^2 \expe{\frac{1}{\Xi_t}\bmat{p_k \\ 1}^\top P_k^{-1}\bmat{p_k \\ 1}}.
\end{align*}
By summing the two sides of the above inequality, we get
\begin{align*}
\expe{\frac{U_t}{\Xi_t}}\leq \mathbb{E}\left[\frac{U_2}{\Xi_t}\right] + \sigma^2\sum_{k=3}^t  \expe{\frac{1}{\Xi_t}\bmat{p_k \\ 1}^\top P_k^{-1}\bmat{p_k \\ 1}}.
\end{align*}
To compute the first term in the above inequality we compute $U_2$. Recall that $p_1$ and $p_2$ are deterministic constants. Then,
\begin{align*}
U_2 &= \frac{1}{(p_2-p_1)^2}\bmat{p_1\varepsilon_1+p_2\varepsilon_2\\ \varepsilon_1+\varepsilon_2}^\top \bmat{2& -(p_1+p_2)\\-(p_1+p_2)&p_1^2+p_2^2} \bmat{p_1\varepsilon_1+p_2\varepsilon_2\\ \varepsilon_1+\varepsilon_2}\\
&=\frac{1}{p_2-p_1}\bmat{p_1\varepsilon_1+p_2\varepsilon_2\\ \varepsilon_1+\varepsilon_2}^\top \bmat{-\varepsilon_1+\varepsilon_2\\p_2\varepsilon_1-p_1\varepsilon_2} \\
&=\varepsilon_1^2+\varepsilon_2^2.
\end{align*}
The desired inequality follows from independence of $\varepsilon_1$ and $\varepsilon_2$ from $\Xi_t$.


\section{Proof of Lemma \ref{lem:1/xi}} \label{app:1/xi}
Let $X_k\sim\mbox{Ber}(p_k)$ and $Y_k\sim\mbox{Ber}(q_k)$ be sequences of independent Bernoulli random variables such that $p_k\geq q_k$ for all $k=1,\ldots,t$. 
Then, for all $c>0$ and all $t\geq 1$ using the law of iterated expectations we get
\begin{align*}
&\mathbb{E}\left[\frac{1}{c+\sum_{k=1}^t X_k}\right] \\
&= (1-p_t)\mathbb{E}\left[\frac{1}{c+\sum_{k=1}^{t-1} X_k}\right] + p_t\mathbb{E}\left[\frac{1}{c+1+\sum_{k=1}^{t-1} X_k}\right] \\
&= \expe{\frac{1}{c+\sum_{k=1}^{t-1} X_k}} - p_t \mathbb{E}\left[\frac{1}{\left(c+\sum_{k=1}^{t-1} X_k\right)\left(1+c+\sum_{k=1}^{t-1} X_k\right)}\right] 
\end{align*}
Using the fact that $p_t\geq q_t$, we get
\begin{align*}
&\mathbb{E}\left[\frac{1}{c+\sum_{k=1}^t X_k}\right] \\
&\leq  \mathbb{E}\left[\frac{1}{c+\sum_{k=1}^{t-1} X_k}\right] - q_t \mathbb{E}\left[\frac{1}{\left(c+\sum_{k=1}^{t-1} X_k\right)\left(1+c+\sum_{k=1}^{t-1} X_k\right)}\right] \\
&=\mathbb{E}\left[\frac{1}{c+\sum_{k=1}^{t-1} X_k + Y_t}\right].
\end{align*}
Using a similar argument, one can show that for all $\tau=1,\ldots,t$, it holds that
\begin{align*}
\mathbb{E}\left[\frac{1}{c+\sum_{k=1}^\tau X_k + \sum_{k=\tau+1}^t Y_k}\right]\leq \mathbb{E}\left[\frac{1}{c+\sum_{k=1}^{\tau-1} X_k + \sum_{k=\tau}^t Y_k}\right].
\end{align*}
By taking the sum of the both sides of the above inequality from $\tau=1$ to $\tau=t$, we get
\begin{align*}
\mathbb{E}\left[\frac{1}{c+\sum_{k=1}^t X_k}\right]\leq \mathbb{E}\left[\frac{1}{c+\sum_{k=1}^t Y_k}\right].
\end{align*} 
Now let $q_k=q$ for all $k=1,\ldots,t$. It follows that in this case $\sum_{k=1}^t Y_k$ is a Binomial random variable with parameters $(t,q)$. Then
\begin{align*}
\mathbb{E}\left[\frac{1}{c+\sum_{k=1}^tY_k}\right]&=\sum_{k=0}^t  \frac{1}{c+k}{t \choose k} q^k(1-q)^{t-k}\\
&=\sum_{k=0}^t  \frac{1+k}{c+k}\frac{1}{1+k} {t \choose k} q^k(1-q)^{t-k}\\
&\leq \frac{1}{c} \sum_{k=0}^t \frac{1}{1+k}{t \choose k} q^k(1-q)^{t-k}\\
&=\frac{1}{c(t+1)q}\sum_{k=0}^t {t+1 \choose k+1}q^{k+1}(1-q)^{t-k}\\
&=\frac{1}{c(t+1)q}\left(1-(1-q)^{t+1}\right).
\end{align*}
Thus,
\begin{align*}
\mathbb{E}\left[\frac{1}{c+\sum_{k=1}^tY_k}\right]\leq \frac{1}{c(t+1)q}.
\end{align*}
It follows from the above that for all $t\geq 3$
\begin{align*}
\mathbb{E}\left[\frac{1}{c+\sum_{k=3}^tY_k}\right]\leq \frac{1}{c(t-1)q}\leq \frac{3}{2ctq}.
\end{align*}
Using the fact that $\eta k^{-r}\geq \eta t^{-r}$ for $k=3,\ldots,t$, and by setting $q=\eta t^{-r}$ we get
\begin{align*}
\expe{\frac{1}{\Xi_t}} = \frac{3}{2\rho^2}\expe{\frac{1}{3(p_2-p_1)^2/(4\rho^2)+ \sum_{k=3}^t\xi_k}}\leq \frac{3}{(p_2-p_1)^2\eta t^{1-r}}.
\end{align*}


\section{Proof of Lemma \ref{lem:errors}} \label{app:errors}
Using the fact that $\xi_t(1-\xi_t)=0$ for all $t\geq 3$, we obtain
\begin{align*}
\mathbb{E}^\gamma\left[(p_t-p^*_t)^2\right]&=\mathbb{E}^\gamma\left[\left((\widehat{p}_t-p^*_t)(1-\xi_t)+(\bar{p}_{t-1}-p^*_t+\rho) \xi_t\right)^2\right]\\
&=\mathbb{E}^\gamma\left[(\widehat{p}_t-p^*_t)^2\right]\mathbb{E}\left[(1-\xi_t)^2\right]+\mathbb{E}^\gamma\left[(\bar{p}_{t-1}-p^*_t+\rho )^2\right]\mathbb{E}\left[\xi_t^2\right]\\
&=\mathbb{E}^\gamma\left[(\widehat{p}_t-p^*_t)^2\right] (1-\eta t^{-r}) + \mathbb{E}^\gamma\left[(\bar{p}_{t-1}-p^*_t+\rho )^2\right] \eta t^{-r}\\
&\leq \mathbb{E}^\gamma\left[(\widehat{p}_t-p^*_t)^2\right] + \mathbb{E}^\gamma\left[(\bar{p}_{t-1}-p^*_t+\rho )^2\right] \eta t^{-r}.
\end{align*}
The second equality follows from the fact that $\xi_t$ is independent from $\widehat{p}_t$ and $\bar{p}_{t-1}$. Using Equation \eqref{pol:myopic p}, it is not difficult to show that 
\begin{align*}
\widehat{p}_t-p^*_t = \frac{1}{2a\widehat{a}_t}\bmat{b\\-a}^\top(\widehat{\theta}_t-\theta).
\end{align*}
Then,
\begin{align*}
|\widehat{p}_t-p^*_t|\leq  \frac{\sqrt{a^2+b^2}}{2a\underline{a}} \| \widehat{\theta}_t-\theta\|.
\end{align*}
Using the above inequality, and the fact that $|\bar{p}_{t-1}-p^*_t+\rho|\leq \overline{p}+\rho$ for all $t\geq 3$, we obtain
\begin{align*}
\mathbb{E}^\gamma\left[(p_t-p^*_t)^2\right]&\leq  k_1 \mathbb{E}^\gamma\left[\|\widehat{\theta}_{t-1}-\theta\|^2\right]+k_2 t^{-r}, 
\end{align*}
where 
\begin{align*}
k_1 &:=\frac{a^2+b^2}{4a^2\underline{a}^2} \quad \text{and} \quad  k_2:= (\overline{p}+\rho)^2\eta.
\end{align*}
Under the randomly perturbed myopic policy for $t\geq 3$, using the law of total expectation we get
\begin{align}
&\hspace{-4em}\mathbb{E}^\gamma\left[(\quant_t-\quant^*_t-a(p_t-p^*_t))^2\right]\nonumber\\
&= (1-\eta t^{-r})\mathbb{E}^\gamma\left[(\widehat{\quant}_t-\quant^*_t-a(\widehat{p}_t-p^*_t))^2\right]\nonumber \\
&\qquad + \eta t^{-r}\mathbb{E}^\gamma\left[(\widehat{\quant}_t-\quant^*_t-a(\bar{p}_{t-1}-p^*_t+\rho))^2\right]\nonumber\\
&\leq \mathbb{E}^\gamma\left[(\widehat{\quant}_t-\quant^*_t-a(\widehat{p}_t-p^*_t))^2\right] + k_6 t^{-r},\label{eq:q-p}
\end{align}
where $k_6$ is defined as
\begin{align*}
k_6:= \eta\left(\frac{1}{2}(\mu^- \overline{a} +\overline{b}-b)+(\overline{\varepsilon}-\underline{\varepsilon})+a\overline{p}\right)^2.
\end{align*}
For the first term in Inequality \eqref{eq:q-p}, using the fact that $\widehat{\quant}_t=\widehat{a}_t \widehat{p}_t + \widehat{b}_t+ \widehat{F}_{t}^{-1}(\alpha_t) $ we get
\begin{align*}
\widehat{\quant}_t-\quant^*_t-a(\widehat{p}_t-p^*_t)&=(\widehat{a}_{t-1}-a)\widehat{p}_t +(\widehat{b}_{t-1}-b) + \widehat{F}_{t-1}^{-1}(\alpha_t)-F^{-1}(\alpha_t)\\
&=\begin{bmatrix} \widehat{p}_t\\ 1 \end{bmatrix}^\top (\widehat{\theta}_{t-1}-\theta) + \widehat{F}_{t-1}^{-1}(\alpha_t)-F^{-1}(\alpha_t),
\end{align*}
Thus,
\begin{align*}
&\hspace{-4em}\mathbb{E}^\gamma\left[(\widehat{\quant}_t-\quant^*_t-a(\widehat{p}_t-p^*_t))^2\right]\\
&=\mathbb{E}^\gamma\left[\left(\begin{bmatrix}\widehat{p}_t\\ 1 \end{bmatrix}^\top(\widehat{\theta}_{t-1}-\theta)+\widehat{F}_{t-1}^{-1}(\alpha_t)-F^{-1}(\alpha_t)\right)^2\right]\\
&\leq \mathbb{E}^\gamma\left[\left(2\sqrt{1+\overline{p}^2} \| \widehat{\theta}_{t-1}-\theta\| + |F_{t-1}^{-1}(\alpha_t)  - F^{-1}(\alpha_t)|\right)^2\right].
\end{align*}
The  inequality follows from Inequality \eqref{eq:quant}. Recall that $\widehat{\theta}_t$ is the TLSE. Thus, it holds that $\| \widehat{\theta}_t-\theta\|\leq \overline{\delta} $ surely, where $\overline{\delta}$ is defined as $\overline{\delta}:=\sqrt{(\overline{a}-\underline{a})^2+\overline{b}^2}$. Thus,
\begin{align*}
&\mathbb{E}^\gamma\left[(\widehat{\quant}_t-\quant^*_t-a(\widehat{p}_t-p^*_t))^2\right]\leq 4(1+\overline{p}^2)\mathbb{E}^\gamma\left[\| \widehat{\theta}_{t-1}-\theta\|_1^2\right] \\
&\hspace{4em}+ \mathbb{E}\left[(F_{t-1}^{-1}(\alpha_t)  - F^{-1}(\alpha_t))^2\right] + 4\overline{\delta}\sqrt{1+\overline{p}^2}  \mathbb{E}\left[|F_{t-1}^{-1}(\alpha_t)  - F^{-1}(\alpha_t)|\right].
\end{align*}
To bound the second and the third terms in the right hand side of the above inequality, we use Inequality \eqref{eq:dvoretzky}, and the fact that $|F_t^{-1}(\alpha_t)-F^{-1}(\alpha_t)|$ and $(F_t^{-1}(\alpha_t)-F^{-1}(\alpha_t))^2$ are non-negative random variables. For $t\geq 2$, we have that
\begin{align}
\mathbb{E}\left[\left|F_{t}^{-1}(\alpha_t)-F^{-1}(\alpha_t)\right|\right]&=\int_0^\infty \mathbb{P}\left\{\left|F_{t}^{-1}(\alpha_t)-F^{-1}(\alpha_t)\right|\geq \delta\right\}d\delta\nonumber\\
&\leq \int_0^\infty 2\exp\left(-\mu_1\delta^2t\right)d\delta\nonumber\\
&=\frac{\sqrt{\pi}}{\sqrt{\mu_1 t}}, \label{eq:abs quantile er}
\end{align}
and 
\begin{align}
\mathbb{E}\left[\left(F_{t}^{-1}(\alpha_t)-F^{-1}(\alpha_t)\right)^2\right]&=\int_0^\infty \mathbb{P}\left\{\left|F_{t}^{-1}(\alpha_t)-F^{-1}(\alpha_t)\right|\geq \sqrt{\delta}\right\}d\delta\nonumber\\
&\leq \int_0^\infty 2\exp\left(-\mu_1\delta t\right)d\delta\nonumber\\
& = \frac{2}{\mu_1  t}. \label{eq:sq quantile er}
\end{align}
We conclude the proof by defining the constants $k_3$, $k_4$, and $k_5$ as 
\begin{align*}
k_3 := 4(1+\overline{p}^2),\quad k_4:= \frac{2}{\mu_1},\quad \text{and} \quad k_5 := 4\overline{\delta}\frac{\sqrt{\pi(1+\overline{p}^2)}}{\sqrt{\mu_1}}.
\end{align*}

\end{appendices}

\end{document}